\documentclass[5p,twocolumn]{elsarticle}        
\usepackage{natbib}
\usepackage{amsfonts,amsmath,amssymb}
\usepackage{graphicx}
\usepackage{subfigure}
\usepackage{bm}
\bibpunct{(}{)}{;}{a}{,}{,}
 \usepackage [english]{babel}
\journal{Ocean Modeling}

\begin{document}

\bibliographystyle{elsarticle-harv.bst}

\begin{frontmatter}
\title{Nearshore Sticky Waters
}
\author[jmr,jmr2,jmr3]{Juan M. Restrepo\corref{cor1}
\cortext[cor1]{Corresponding Author.}}
\ead{restrepo@math.arizona.edu, Ph. 520-621-4367 }
\author[jmr]{Shankar C. Venkataramani}
\author[cd1,cd2]{Clint Dawson}
\address[jmr]{Department of Mathematics, University of Arizona, Tucson, AZ 85721 U.S.A.}
\address[jmr2]{Department of Atmospheric Sciences, University of Arizona, Tucson, AZ 85721 U.S.A.}
\address[jmr3]{Department of Physics, University of Arizona, Tucson, AZ 85721 U.S.A.}
\address[cd1]{Department of Aerospace Engineering and Engineering Mechanics, University of Texas, Austin TX 78712, U.S.A.}
\address[cd2]{Institute of Computational and Engineering Sciences, University of Texas, Austin TX 78712, U.S.A.
}

\begin{abstract}
Wind- and current-driven  flotsam, oil spills, pollutants, and
nutrients,  approaching the nearshore will frequently
appear to slow down/park just beyond the break zone, where waves break.
Moreover, 
the portion of these tracers that beach will do so only after a long
time. Explaining why these tracers park and at what rate they reach the shore
 has important implications on
a variety of different nearshore environmental issues, including the determination of  what  subscale processes are essential in  computer models for
the simulation of pollutant transport in the nearshore.  Using a simple model 
 we provide an explanation for the underlying mechanism responsible
for the  parking of tracers, not subject to inertial effects, 
the role played by the bottom topography, and 
the non-uniform dispersion which leads, in some circumstances,  to the eventual landing of 
all or a portion of the tracers.  We refer to the parking  phenomenon in this environment as  nearshore sticky waters.
\end{abstract}
\end{frontmatter}
\section{Introduction}
\label{intro}

\begin{figure*}
\centering
\includegraphics[width=0.75 \textwidth]{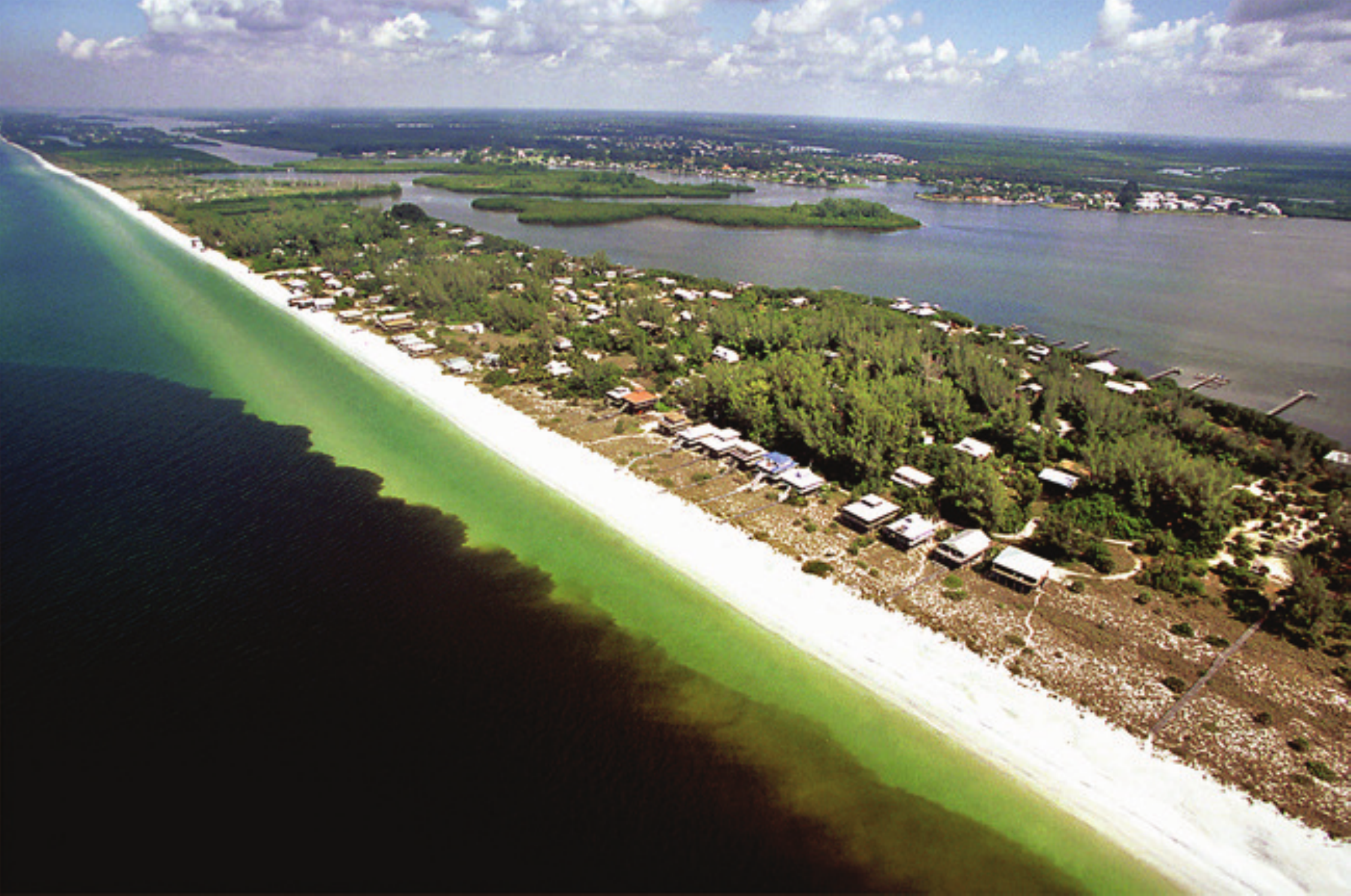}
\caption{A red tide event, off the coast of Florida. The event occurs nearly annually along the state's Gulf Coast.
 Image courtesy of P. Schmidt, Charlotte  Sun. For an example that has more surfzone wave action, see Figure 1 of \cite{grant05}.}
 \label{red}
\end{figure*}

Oil from spills,  red tides, flotsam and other suspended and surface tracers
approach the nearshore, carried by winds and currents. It is not uncommon, however, 
that these debris and tracers  slow down  and park themselves, somewhere beyond the break zone ( See Figure~\ref{red}); 
eventually, a portion of these reach the beach zone by the action of turbulence and tidal effects, in
combination with inertial effects on  the debris.
The tendency of tracers to park themselves
in certain areas of the Great Barrier Reef has been noted. \cite{wolanski}, who reported the phenomenon, and denoted it as  ``sticky waters." Though we will not be discussing estuarine environments and the mechanism at play in the Great
Barrier Reef situation may be different from the nearshore case, we will borrow this terminology and refer to the phenomenon we investigate in this paper as ``nearshore sticky waters."

Of obvious environmental, economic, and social importance, 
understanding why nearshore sticky waters occur is also fundamental to improved  environmental assessments of  coastal settings.
Moreover, as part of a larger research agenda
aimed at improving models for pollutant transport in ocean general circulation models, 
nearshore sticky waters offers a field-verifiable problem with which to test contaminant advection 
reaction and  dispersion models.

The focus is on  tracer transport phenomena, with length scales several times larger than the depth and temporal scales of hours, weeks. That is, we are mostly concerned with large-scale pollution ``disasters," such as large-scale red tides, significant oil spills, etc.
Although we consider long time and space scales, we cannot ignore depth dependent features of the flow and the transport of tracers. The tracer may be buoyant but not necessarily entirely residing on the surface of the ocean; We therefore consider a layered (instead of simply depth averaged) model for the tracer and account for the vertical structure of the advective velocity. We defer consideration of tracers with non-trivial inertial effects to a separate study.  Obviously, tracers advect and diffuse in the alongshore direction as well as in the cross-shore direction. In fact, advection/diffusion in the longshore direction is usually more intense in many non-estuarine  environments. However, if we consider a situation where the longshore variations of the tracer concentrations are small, the divergence of the flux in the longshore direction is negligible, and it is appropriate to consider a one-dimensional problem in the cross-shore direction.

Nearshore sticky waters will refer to the slowing down or the parking of the tracer approaching the shore. In a sticky water situation the center of mass of the incoming tracer that is approaching the shore at advective speeds will experience a partial or total slowing down. Whatever tracer amounts reach the shore will do so by the action of dispersive effects, usually higher inside the breakzone than in deeper waters. In the cross-shore direction, large scale currents are typically weak, close to the shore. In a wave-dominated nearshore setting, the typical advection velocity would be the residual flow due to the waves, the Stokes drift velocity. The length scales are those of the long waves, {\it i.e.}, waves which have wavelengths that are large when compared to the depth. The diffusive length scale is typified by large-scale eddies; if the break zone is a significant source of mixing, the length scale would be the distance between the start of the breaking of the waves and the shore. When advective and diffusive effects are those in balance, the diffusive time scale is large, in the order of hours.
There is consensus that in wave-dominated beaches the dissipation of waves is different inside and outside of the breakzone, the latter being considerably smaller than the former. \cite{meibook},  Chapter 10, describes the theoretical development of a model for wave action dissipation, based upon dimensional analysis and homogeneous turbulence concepts (see also \cite{svendsenpet}, for further developments). The model used in \cite{uchiyamalongshore} is that of \cite{TG83}, which is one of several based upon hydraulic jump parametrizations. Analysis of field data of the dispersion of tracers in the nearshore suggest that the diffusivity is much higher in the break zone than outside. Dispersion estimates based upon the dimensional analysis model of \cite{svendsenpet} are off by orders of magnitude, when compared to field data (see 
\cite{feddersenjpo12}). 
A possible explanation for the discrepancy might lie in the fact that the dimensional parametrization is 
based upon homogeneous turbulence conditions and is more typical of the smaller scale vertical diffusion, rather than the larger eddy-scale transverse diffusion (Feddersen, private communication).

The basic depth-averaged hydrodynamics, appropriate to these scales, 
 are captured by the similarly scaled vortex force model in
\cite{MRL04} (see \cite{MR99, r01} for background and \cite{LRM06} for a comparison of this ``vortex force'' model and the ``radiation stress'' alternative. See also \cite{smithjpo06}). In the following form, the model has been used to study nearshore problems, such as longshore currents (in \cite{uchiyamalongshore}), and rip currents
(in \cite{weirrips}). The depth-averaged momentum balance reads:
\begin{equation}
\frac{D\mathbf{v}^c}{D t}=-g\nabla\zeta  - \chi  [\mathbf{u}^{St}]^\bot  + {\bf N},
\label{mom}
\end{equation}
where $\chi$ is the vorticity of the depth-averaged velocity ${\bf v}^c(x,y,t):=(u^c,v^c)$, and ${\bf N}$ encompasses bottom drag, wind forcing, and dissipation; it also encompasses momentum transfers from wave breaking to the current momentum (see \cite{RRMB}). The vortex force is the second term on the right hand side, which couples the residual flow due to the waves to the rotation in the current ${\bf v}^c$. The depth-averaged Stokes drift velocity is denoted by $\mathbf{u}^{St}:=(u^{St},v^{St})$;  the operator $\bot$ is used to obtain $[\mathbf{u}^{St}]^\bot = (-v^{St},u^{St})$.   

The continuity equation reads 
\begin{equation}
\frac{\partial \zeta}{\partial t}= -\frac{\partial \zeta^c}{\partial t}-\nabla \cdot [{\cal H} ({\bf v}^c +  \mathbf{u}^{St})],
\label{continuity}
\end{equation}
where ${\cal H} = \zeta^c + H(x,t)$ is the local water column depth and $\zeta^c = \zeta + \hat \zeta$ is the composite sea elevation; $\hat \zeta$ is the quasi-static sea elevation. The waves are found  via conservation equations  for the wave action ${\cal A}$, and wavenumber ${\bf k}$.  For the wave action, the equation is
\begin{equation}
\frac{\partial {\cal A}}{\partial t} + \nabla \cdot ({\bf C}_G {\cal A}) = N_{{\cal A}},
\label{waveevo}
\end{equation}
where $N_{{\cal A}}$ is the loss term and
${\bf C}_G$ is the absolute group velocity,
\begin{equation}
{\bf C}_G = {\bf v}^c + \frac{\Sigma}{2 k^2} \left(1 + \frac{2 k {\cal H}}{\sinh 2 k {\cal H}} \right) {\bf k}.
\label{groupvelo}
\end{equation}
 The relative frequency is $\omega = {\bf v}^c\cdot{\bf k} + \Sigma$, where 
 the frequency  satisfies the dispersion relation $\Sigma = \sqrt{g k \tanh (k {\cal H})}$.
The wave action,  the Stokes drift velocity and the quasi-static sea elevation response are given by
\begin{equation}
{\cal A}:=\frac{1}{2\Sigma}\rho g A^{2}, \quad \mathbf{u}^{St}:=\frac{1}{\rho {\cal H}}{\cal A}\mathbf{k}, \quad \hat \zeta =-\frac{A^2 k}{2 \sinh(2 k {\cal H})},
\label{waveac}
\end{equation}
respectively. $A$ is the wave amplitude and $k$ is the magnitude of the wavenumber ${\bf k}$.
 The wavenumber conservation equations are
\begin{equation}
\frac{\partial {\bf k}}{\partial t} + \nabla (\Sigma +{\bf v}^c \cdot {\bf k})
 =0.
\label{wavenumber}
\end{equation}

The evolution equation for a  tracer $\theta$,  (see \cite{MRL04}), is
\begin{equation}
\frac{\partial \theta}{\partial t} + ({\bf v}^c + {\bf u}^{St})\cdot \nabla \theta =
N_{\theta},
\label{tracer}
\end{equation}
where $N_\theta$ is the tracer dispersion term.

The simplest situation we consider is that of a flow with mean shoreward-directed velocity, transporting the pollutant toward land, flowing over  a sloped and featureless bathymetry.  
We consider a nearshore domain that has only transverse extent $x$ and  depth $z$; the water column increases  in depth, away from the shore. Consideration of the actual mechanism that is generating the current
field makes the basic story presented here richer, but is beyond the scope of this paper. Instead we focus on  the basic kinematics of the tracers. 

The advecting mean current, with a shore-directed component,  might consist entirely or partially 
of a wave-induced flow, the Stokes drift velocity (see \cite{meibook}).  For specificity we will assume, in fact,  that the advective mean current is exclusively composed of the Stokes drift and that these are generated by shore-directed waves.
(As the reader will eventually surmise  we could have assumed instead the presence of currents not associated with waves, or even considered the case where both wave-induced flows and  currents are present; nearshore sticky waters conditions do not require the presence of wave-generated currents). According to (\ref{mom}),  however, this Stokes drift will not generate a vortex force. If the  velocity at the shore end, at $x=0$, is zero, the cross-shore component of the depth-averaged current $u^c(x,t)$ must be equal and opposite to  
the cross-shore component of the depth-averaged Stokes drift velocity   (in \cite{uchiyamalongshore} we recognized it as the {\it anti-Stokes} current, but more generally it is the undertow current. See \citet{lentzfewings} for more details concerning the anti-Stokes current. This is a very readable introduction to the nearshore flow environment).  The Stokes drift velocity depends on 
the wave action and the wavenumber  by (\ref{waveac}) and (\ref{wavenumber}).  In \cite{uchiyamalongshore}, Figure 1, are shown the somewhat typical slight increases in the wave action, as the waves approach the breakzone, followed by their partial or full dissipation due to 
wave breaking, with an ensuing transfer of momentum to the currents. The resulting cross-shore component of the Stokes drift velocity increases as the waves shoal, and then diminishes drastically or becomes insignificant.  If a separate current can be identified in the flow, and waves are present, an ensuing transfer of momentum ensues when the waves break. The currents are also subjected to 
bottom/form drag, sea elevation gradients, etc. 

Momentum transfers from the breaking  wave field to the currents via ${\bf N}$ in (\ref{mom}),
can generate currents, however, we will assume in what follows that velocities thus so generated are 
inconsequential.  The depth-dependence of the current will prove essential:
 The Stokes drift drops off exponentially as a function of the depth. The velocity at $z=-H(x)$, 
 must satisfy $W=0, {\bf U}:=(U,V) = 0$, where $W$ and ${\bf U}:=(U,V)$ are the vertical and transverse three-dimensional and time dependent Eulerian velocity components, respectively.
(The boundary condition $W=-{\bf U}\cdot \nabla H$,  applies to  an inviscid flow).  
We assume that the wavenumber ${\bf k} = -k \hat x$, where $\hat x$ is the unit transverse vector, pointing away from the beach. With this assumption, and   (\ref{waveac})-(\ref{groupvelo}) it is understood that  reflections from the shore are insignificant, thanks to dissipative processes embodied by $N_{\cal  A}$.

Since ${\bf v}^c + {\bf u}^{St}=0$ near the beach, the evolution of the depth-averaged tracer there  is purely diffusive, 
 by (\ref{tracer}). It is generally agreed that diffusivity is small outside of the breakzone. If the diffusivity  near the shore were insignificant,  
 there would be little change of the depth-averaged tracer distribution.  
 However, the flow in the breakzone is complex. At the larger scales these turbulent and boundary layer effects manifest themselves as enhanced diffusivity and viscosity. 
   
\section{The Model} \label{sec:pde}

In this paper, we propose a {\em kinematic} model for oil transport in the nearshore, and in the model we include the following effects (1) A mean advective flow that is depth dependent and is shore directed on the surface, (2) A dispersion model that models the transport due to the fluctuating component of the velocity, and accounts for the enhanced diffusion in the break zone, (3) A simplified oil model which includes the effects of buoyant stratification of the oil into a surface slick and a bulk suspension, and (4) An exchange interaction which allows the bulk suspended oil to resurface and turbulence to entrain the surface oil into the bulk. The model is purely kinematic, and applies for various ``physical" models of the underlying mechanisms with appropriate parameterizations of the advective velocity, dispersion, mixing layer depth and exchange rate.

\begin{figure}
\centering
\includegraphics[width= \linewidth]{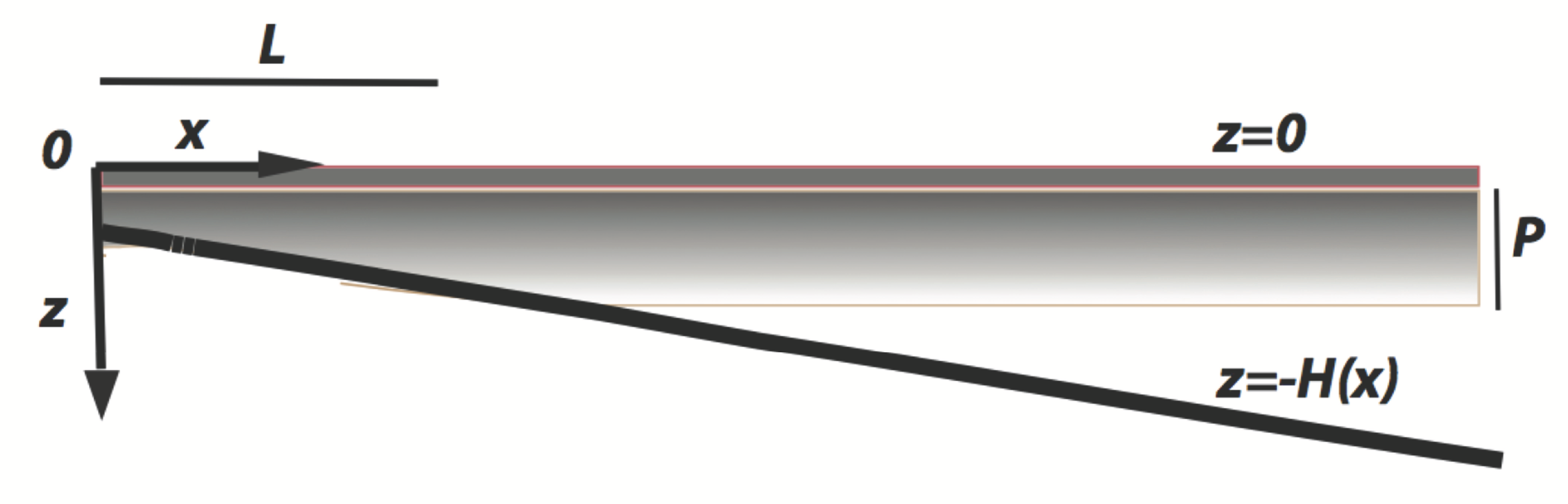}
\caption{Schematic cross-section of the model domain. A light, thin oil slick sits atop the ocean. The
ocean's mixed layer of thickness $P$ is laden with oil droplets, accounted for as a concentration. The distance from
the shore, at $x=0$, is denoted by $x$.  The break zone extends to $x=L$. The ocean surface is at $z=0$ and bottom topography is fixed and described by $z=-H(x)$.}
\label{geo}
\end{figure}

Figure~\ref{geo} depicts the physical domain. 
The quiescent ocean level is at $z=0$, the basin is bounded below, at $z=-H(x)$.
The domain extends from $x=0$, the shore end, where the depth is $H_0 \ge 0$, to    $x=X$
where the depth is $H_\infty \ge H_0$. 
The bathymetry $H(x)$  will be  sloped and featureless:
\[
H(x) = 
 H_0 +m x,  \quad 0 \le x \le X, 
  \]
  where $H_0$ is the depth in the nearshore, and $m \ge 0 $ is the slope.
We distinguish two oceanic regimes in our problem: the high mixing surf zone,
corresponding to  $0 \le x \le L$,  and the deep ocean zone, from $L <x \le X$. 
$L$ is  typically tens to hundreds of meters. The
pollutant (for example, oil), or the tracer (for example, an algal bloom)
is  subject to buoyancy effects. Oil in the surface slick may be entrained by the action of wave breaking and turbulent mixing. The oil may also resurface, at a rate dependent on the size of the droplets.
We will assume that, in the most general case, there is a very thin layer of pure oil, riding on the ocean surface. This layer, which we will denote as the {\it oil slick}, has thickness $s(x,t)$, typically micrometric. Immediately below is a layer of ocean in which the bulk of the oil  is found, in suspension.  As depicted in Figure~\ref{geo}, the layer containing the suspended oil is assumed to have a maximum  thickness $P$.  (It is possible to estimate $P$ in terms of the flow, but doing so is beyond the scope of this study. Here we treat $P$ as a parameter). We will denote this oil in suspension 
as the {\it interior oil}. $b(x,t)$ is thickness of an ``equivalent"  pure oil layer containing the same amount of oil as the interior. Assuming that the interior oil is uniformly distributed within the mixed layer, we  have the {\em equation of state}
\begin{equation}
 b(x,t) =B(x,t) \xi(x),
 \label{eqstate}
 \end{equation}
 where $B$ denotes the (dimensionless) volume fraction of the oil in suspension, and $\xi(x)$ is the local depth of the mixed layer.
  We approximate $\xi(x)$ as a smooth approximation to $\min{(H(x), P)}$. 

\begin{figure*}
\begin{center}
\subfigure[]
{
\includegraphics[height = 0.23 \textwidth]{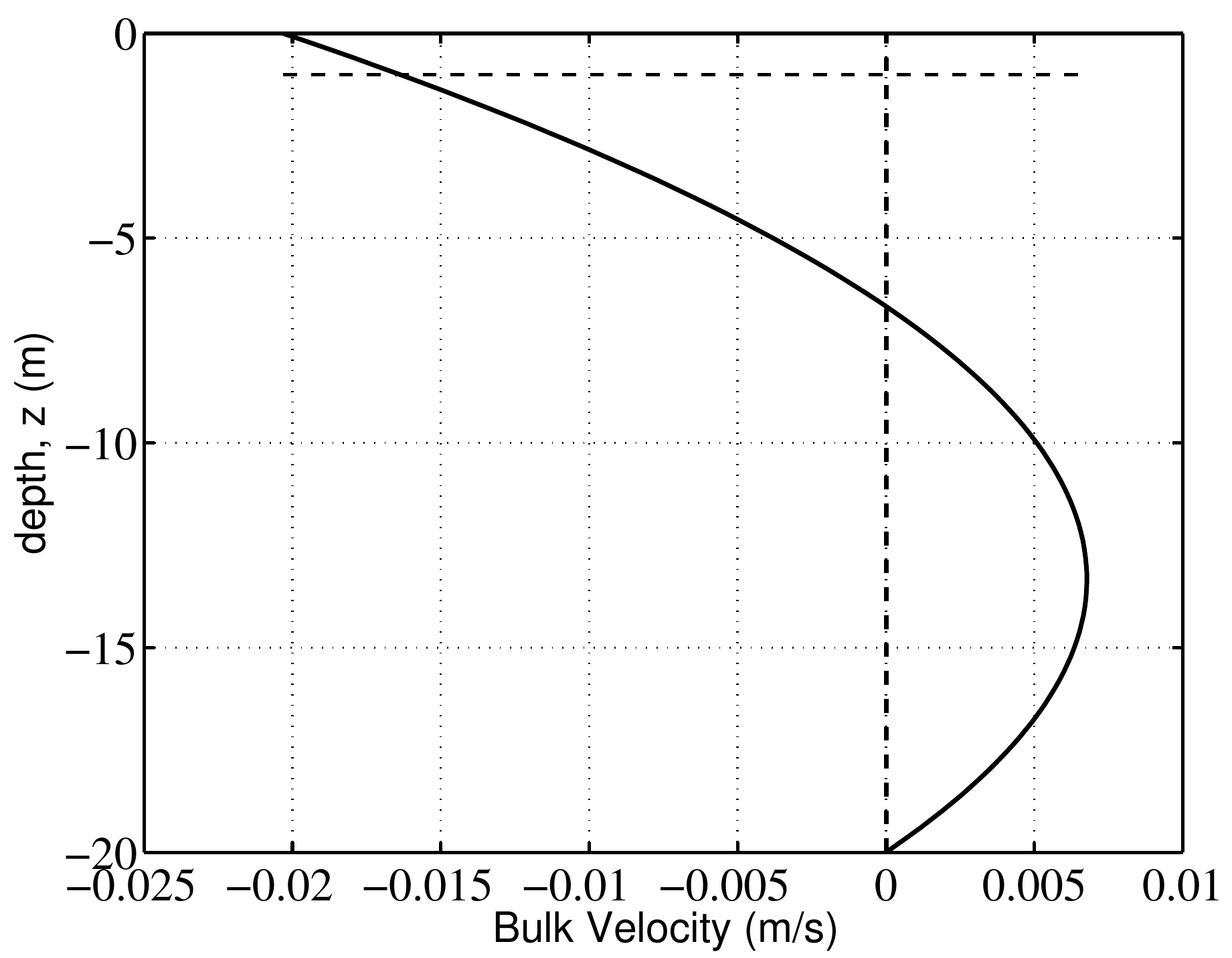}
\label{veloa}
}
\subfigure[]
{
\includegraphics[height = 0.24 \textwidth]{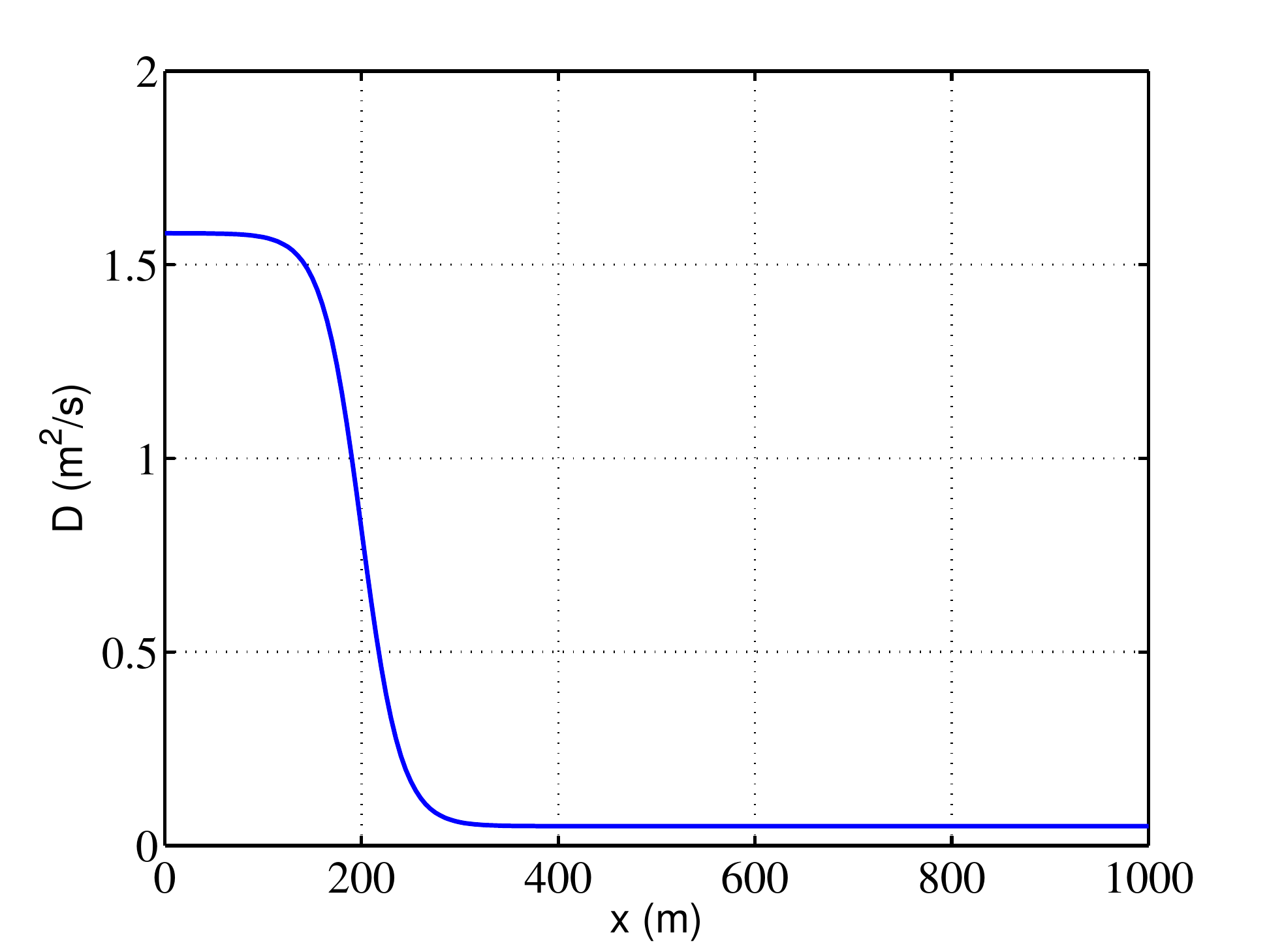}
\label{velob}
}
\subfigure[]
{
\includegraphics[height = 0.23 \textwidth]{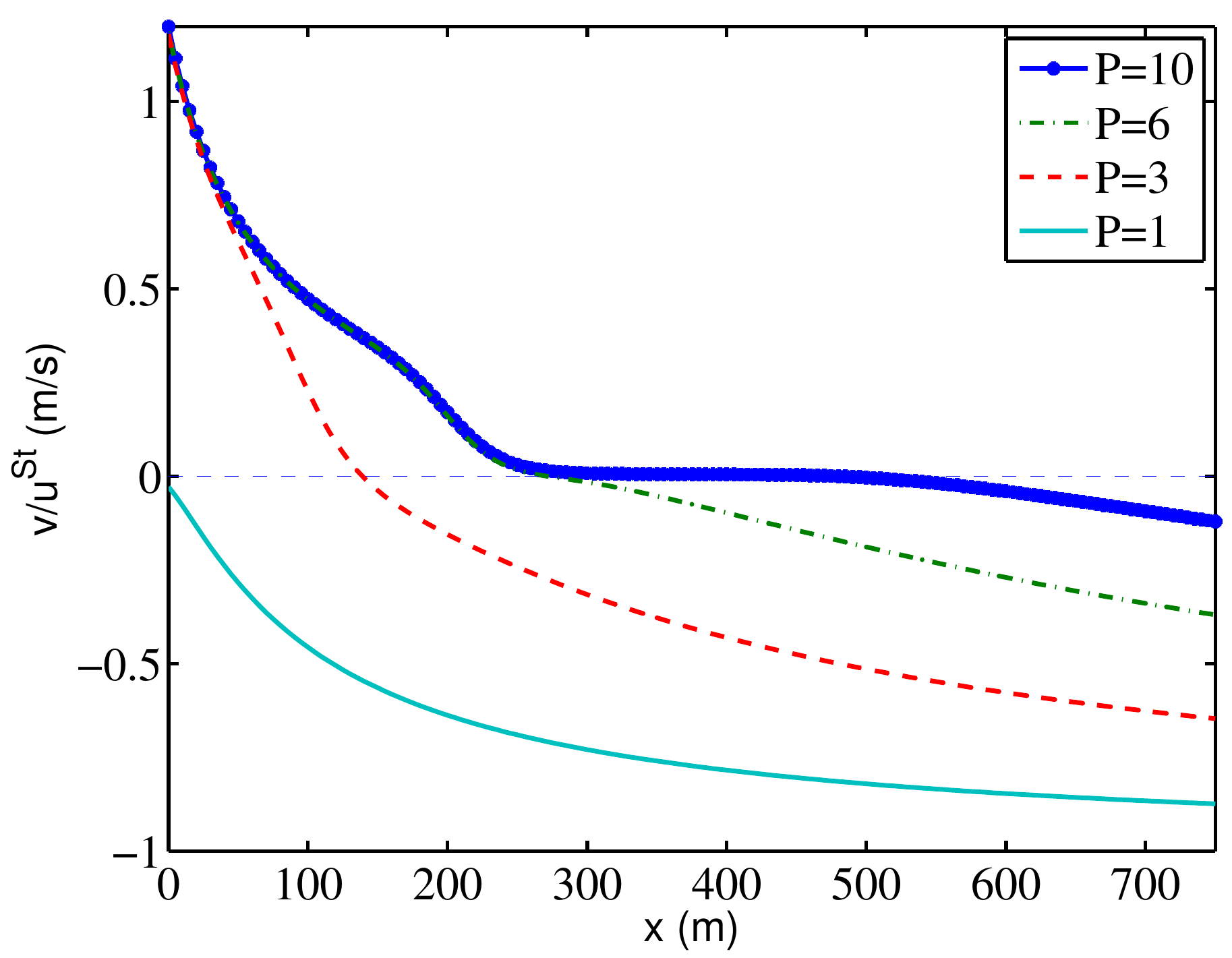}
\label{veloc}
}
\caption{\label{velo}For the examples considered, \subref{veloa} parabolic  velocity profile $u(x,z)$, at $H=20$m;
\subref{velob} dispersion $D(x)$, over the whole domain. It increases substantially close to the shore, where the flow is more  turbulent. \subref{veloc} The normalized effective velocity $v(x)/|{\cal U}^{St}|$ in \protect{\eqref{altoilb}},  for different mixing depths $P$.
See Table~\protect{\ref{tab1}} for parameters.}
\end{center}
\end{figure*}

We now discuss the various mechanisms that transport the tracer (oil) and build a mathematical model for this process.   Because the oil slick moves with the surface, the cross-shore component of the oil slick velocity  $u_{S} \approx U^{St}(x,0,t):={\cal U}^{St}$, the Stokes drift velocity, evaluated at the surface, ${\cal U}^{St} \approx -A^2 k \Sigma$. 
(The mean Eulerian velocity has been set to zero). More realistically Stokes drift velocity depends on $x$ since the waves obey a 
 dispersion relation that is a function of $x$, via its dependence on the water column depth; however, the results presented in the analysis that follows would only change quantitatively.
 (Idealized descriptions for simple shore geometries and waves can be found in \cite{lentzfewings}).
The (time-averaged) advective flux in the slick is thus given by $u_S s(x,t)$. We will be using a simple proxy in place of the actual velocity $U(x,z,t)$, consistent with
the essential kinematics detailed in Section \ref{intro}: Assuming a parabolic profile for the Lagrangian mean velocity (See Figure~\ref{veloa}), we get 
 $\displaystyle{U={\cal U}^{St}\left( 1+\frac{4 z}{H(x)} +\frac{3 z^2}{H(x)^2}\right)}$ by requiring that   $U$  equals ${\cal U}^{St}$ at $z=0$, equals zero at $z= -H(x)$ and has zero depth average {\em i.e.}, $\int_{-H}^0 U \, dz = 0$. Undoubtedly, this is a crude approximation of the real flow, consisting of a 
 flow, with mean shoreward-directed velocity, and an undertow, but it is qualitatively consistent with \cite{pearson2009}, Figure 8, (ignoring mass fluxes that shift the mean sea level away from $z=0$, {\it cf.}, (\ref{waveac}) ). Averaging $U$  over the mixed layer $-\xi(x) \leq z \leq 0$, we get the {\em bulk velocity} 
 \begin{equation}
 u_B(x) ={\cal  U}^{St} \left[H(x) -  \xi(x)\right]^2/H(x)^2. 
 \label{uvel}
 \end{equation}
The (time-averaged) advective flux in the bulk is thus given by $u_B(x) b(x,t)$.

The velocity fluctuations about the mean contribute to an effective dispersion (diffusivity) of the tracer. Field measurements indicate that the vertical and horizontal diffusivities have different dominant mechanics and different magnitudes (see \cite{feddersenjpo12}, \cite{feddersenjgl12}
and references cited in these). At scales of the surfzone itself, which are much larger, typically, than the water column depth, the horizontal diffusivity is dominated by eddies at scales larger than the depth. Measurement, characterization and parametrizations of the diffusivity tensor in the nearshore and the surfzone has come a long way in the last 20 years.  Recent field measurements of large scale cross shore and long shore diffusivities are reported in \cite{clarketal}. These appear to be poorly captured in every respect by older estimates derived by  dimensional analysis arguments in \cite{svendsenpet} (see also \cite{pearson2009})
with regard to cross-shore diffusivities, for example. This is clear in Figure 13 of \cite{clarketal}. 

For the purpose of formulating a simple model for the sticky water phenomenon we will adopt a
very crude cross shore dispersion model.
From field campaigns, \cite{spydell} give 
a rough estimate of the long-term  cross shore diffusivity in the bulk, of $0.05$ m$^2$/s, away from the break zone, and $0.5-1.75$ m$^2$/s in the nearshore, under mild sea conditions. (Noted in their report, however, is that the diffusivity changes over time and it is generally larger in the longshore direction). 
We assume that the interior oil  and the  slick viscosities are equal,  and given
by the simple model
\begin{equation}
D(x) = D_{eddy} + {\cal S}(x) D_{L},
\label{diffuse}
\end{equation}
where ${\cal S}(x) = (1+\exp[(x-L)/w])^{-1}$. This crude model is similar to the one
proposed in \cite{rippy13}, which is inspired by field measurements.  In the examples that follow $L = 200$m  and $w=20$m  is the width of the transition of the sigmoid. $D_{eddy}=0.05$m$^2$/s is the background eddy diffusivity,  and $D_{L} = 1.6$ m$^2$/s the enhanced diffusivity in the   nearshore due to  turbulence and wave breaking (Figure~\ref{velob}). Fick's law gives a diffusive flux $\displaystyle{-D(x) \frac{\partial}{\partial x} s}$ on the surface. In the subsurface region, the diffusive flux is driven by gradients of the bulk concentration $B$ but its effect on $b$ is found via the equation of state, (\ref{eqstate}). Fick's law along with integration in the depth of the mixed layer gives a diffusive flux $\displaystyle{-\xi(x) D(x) \frac{\partial}{\partial x} B}$ in the bulk.

The last process we model is the exchange of material between the surface and the bulk due to wave-mixing and buoyancy. On long time scales corresponding to averaging over many waves, a simple model for the net flux from the slick into the suspension is the linear expression $ \frac{1}{\tau(x)}((1-\gamma) s - \gamma P B)$, where $\gamma \in (0,1)$ is a parameter which sets the relative proportions of the oil in the slick and in suspension for vertical equilibrium, {\em viz.}, $$
\frac{s}{PB} = \frac{ \gamma}{1-\gamma}. 
$$
$\tau(x)$ is the time scale for the vertical mixing. It is a measure of the total kinetic energy (TKE) in the fluctuating velocity field. Consequently dimensional analysis suggests that $\tau(x)  \approx P^2/D(x)$, the diffusion time scale over the typical depth of the mixed layer.  (See \cite{tkalichbreak}, and references contained therein, for an alternative formulation of this exchange flux  and for more details on the mixing/buoyancy physics).

Neglecting sources and sinks of oil, the conservation laws for $s$ and $b$, obtained from the above fluxes are:
\begin{align}
\frac{\partial s}{\partial t} +  \frac{\partial [u_{S}(x) s] }{\partial x} & = -\frac{(1-\gamma) s - \gamma P B}{\tau(x)} + \frac{\partial}{\partial x}\left[D(x) \frac{\partial s}{\partial x} \right],
\label{oilsl} \\
\frac{\partial b}{\partial t} +  \frac{\partial[u_B(x)  b]}{\partial x} & =  \frac{(1-\gamma) s - \gamma P B}{\tau(x)}  \\
& + \frac{\partial}{\partial x} \left[ \xi(x)  D(x) \frac{\partial}{\partial x}  B\right], \nonumber \\
\implies \frac{\partial b}{\partial t} +  \frac{\partial[v(x)  b]}{\partial x}& = \frac{(1-\gamma) s - \gamma P B}{\tau(x)}  + \frac{\partial}{\partial x} \left[ D(x) \frac{\partial b}{\partial x} \right], \quad
\label{altoilb}
\end{align}
where we have used \eqref{eqstate} and 
\[
v(x):= u_B(x) + D(x) \frac{1}{\xi(x)} \frac{d \xi(x)}{dx},
\]
 (see Figure ~\ref{veloc}). Note that the ``effective'' velocity $v$ of the oil in suspension is not the depth-averaged velocity of the flow $u_B$; rather it has a contribution that depends on the eddy diffusivity $D(x)$ and also the bottom topography $H(x)$ if  $H(x) < P$, the maximum depth of the mixed layer. 

To conserve total mass we specify zero flux conditions at $x=0$ and $x = L$:
\begin{align}
u_S(x) s - D(x) \frac{\partial}{\partial x} s & = 0 \mbox{ at } x=0 \mbox{ and }L, \nonumber \\
 v(x) b - D(x) \frac{\partial}{\partial x} b & = 0 \mbox{ at } x=0 \mbox{ and } L.
\label{no-flux-bcs}
\end{align}
Eqs.~\eqref{oilsl},\eqref{altoilb}~and~\eqref{no-flux-bcs}, together with specified initial conditions $s(x,t=0)$ and $b(x,t=0)$ give a complete mathematical model which can be solved numerically. We denote this model as the {\it PDE model}.

\section{Finite dimensional reduction of the model} \label{sec:ode}

By specifying appropriate functional forms for the diffusivity $D(x)$ and the vertical mixing time $\tau(x)$ in the PDE model Eqs.~\eqref{oilsl},\eqref{altoilb}~and~\eqref{no-flux-bcs}, we can describe a variety of scenarios. Conversely, a natural question is the extent to which simple choices for the functions, $D(x)$ and $\tau(x)$, as motivated in the previous section, yield results that are valid for real physical flows. In this section we demonstrate that the results we obtain from the PDE model are robust, and depend only on gross features of $D(x)$ and $\tau(x)$, but not on fine details of these functions.

To this end we first derive a reduced finite dimensional model that  describes the evolution of Gaussian pulse initial conditions. The ratio of the vertical and horizontal time scales is given by $(P^2/D)  ({\cal U}^{St}/X) \lesssim 0.01$ for $P \sim 1$ to 6m, even in the deep ocean where $D$ is small.  Thus, the vertical tracer distribution  equilibrates very rapidly. We can determine $s$ and $b$ in terms of  the total oil  $q=s+b$   by setting the exchange flux to zero yielding 
$$
 s \approx \frac{\gamma P}{\gamma P + (1-\gamma) \xi} q, \qquad  b \approx \frac{(1-\gamma) \xi}{\gamma P + (1-\gamma) \xi} q.
$$
Substituting in the equations for $s$ and $b$, we obtain
\begin{align}
\frac{\partial q}{\partial t} & = \frac{\partial}{\partial x}\left[D(x) \frac{\partial q}{\partial x} - u_{e}(x) q\right], \nonumber \\ 
u_{e}(x) & = \frac{\gamma P u_{S} + (1-\gamma) \xi v(x)}{\gamma P + (1-\gamma) \xi}.
\label{totaloil}
\end{align}
Observe that these equations no longer depend on $\tau(x)$, showing that as long as the mixing in the vertical direction is rapid on the scale of the horizontal transport, the precise choice of the mixing time is irrelevant.

\subsection{Initial asymptotics: Pulse solutions}

Setting
$\displaystyle{q \approx \frac{1}{\sqrt{2 \pi \sigma^2(t)} }\exp\left[ - \frac{(x-\mu(t))^2}{2 \sigma^2(t)}\right]}$ describes an evolving unit mass Gaussian pulse with mean (peak location) $\mu(t)$ and variance (width)  $\sigma^2(t)$. We expect this Gaussian pulse {\em ansatz} to be valid as long as the peak of the pulse is ``far" from the shore on the scale of its width,  $\mu(t) \gg \sigma(t)$. Using this ansatz in  \eqref{totaloil}  and computing the first and second  moments yields the
{\it ODE model}, 
\begin{equation}
\frac{\partial}{\partial t} \mu = u_{e}(\mu) + \frac{\partial D}{\partial x} (\mu), \qquad
\frac{\partial}{\partial t} \sigma^2  = 2 D(\mu),
\label{initial-asymptotics}
\end{equation}
where we have assumed that  $D$ and $u_{e}$ are slowly varying on the scale of the width $\sigma(t)$.
This model works best at early times, when $\sigma(t)/\mu(t) \ll 1$, so the pulse does not ``feel" the  boundary condition at the shore $x=0$.

\subsection{Steady states}

 Solving  \eqref{totaloil} for long times $t \to \infty$, we obtain the {\em steady state} 
\begin{equation}
q \to q_{\infty} = C \exp\left[\int \frac{u_e(x)}{D(x)} dx\right], C \mbox{ is a normalizing constant}. 
\label{steady}
\end{equation}
$q_\infty$ has a maximum at the $x$ values where $u_e$ changes sign.  Outside the break zone ($D(x)$ is small), or in sufficiently deep water ($\xi(x) = P$ so $\xi'(x) = 0$), we have   $v(x) = u_B(x) < 0$. Thus  $u_e$ is negative (shoreward). 

If $u_e$ from \eqref{totaloil} is negative for  all $x$ the maximum of the steady state distribution of $q$ is at $x=0$, and this is the case if $H_0$ is sufficiently large (see Figure~\ref{veloc}). However if $P > H_0$, then the depth averaged bulk velocity $u_B$ is zero at $x = 0$. In this case, $v(0) = D(0) m/H_0 > 0$ so the  effective velocity $u_e > 0$ for sufficiently small $\gamma$ (see \eqref{totaloil}). In this scenario $u_e$ will change sign away from $x=0$ (see Figure~\ref{veloc}).

Of course, in a physical situation, we do not get to choose $\gamma$. A natural question is in what circumstances is the maximum of the steady state distributions ``significantly" away from the shore, {\em i.e.} the location of the maximum is on the scale of $L$, the width of the break zone. As we will see below, this corresponds to a significant slowing of the tracers as they approach the shore. The following argument gives estimates of the relevant parameter regime. 

From \eqref{uvel}, we see that $u_B = 0$ for all $x <  C = (P-H_0)/m$. This motivates the definition of the non dimensional parameter 
\begin{equation}
 \beta = \frac{C}{L} = \left(\frac{P-H_0}{H_\infty-H_0} \right) \frac{X}{L}.
\label{beta}
\end{equation}
If $\beta < 0$ then  $P < H(x)$ for all $x$. If $0\leq \beta \leq 1$, the point where $P=H(x)$ is in the break zone, and if $\beta > 1$, this point is outside the break zone. 

For $x < C$, i.e. the region where $H(x) < P$, $u_B = 0$ and the definition of $v$ in \eqref{totaloil} yields
$$
u_e(x) > 0 \Leftrightarrow  (1-\gamma) D(x) m \geq -\gamma P {\cal U}^{St} = \gamma P |{\cal U}^{St}|
$$
where we note that the stokes drift ${\cal U}^{St}$ is negative as it is directed shoreward. This condition can be rearranged to give
\begin{equation}
D(x) > \left(\frac{\gamma}{1-\gamma}\right) \frac{P |{\cal U}^{St}|}{m} \equiv D_{threshold}.
\label{threshold}
\end{equation}
Using typical values for $P, {\cal U}^{St}$ and $m$, we get $\frac{P |{\cal U}^{St}|}{m} \sim (0.25-5)$ m$^2$/s so unless $\gamma$ is incredibly small, $D_{threshold} \gg D_{eddy}$ the eddy diffusivity outside the break zone, justfying our claim above that  $u_e < 0$ (shoreward) outside the break zone. To get that $u_e(x) > 0$ for $x$ on the scale of $L$, we thus need two conditions to hold. From \eqref{beta}, we need $\beta \gtrsim O(1)$, and from \eqref{threshold} and \eqref{diffuse}, we need that $D_L \geq D_{threshold}$. This motivates the definition of a second dimensionless parameter 
\begin{equation}
\delta = \frac{D_L}{D_{threshold}} = \frac{(1-\gamma) D_L (H_\infty - H_0)}{\gamma P |{\cal U}^{St}| X}.
\label{delta}
\end{equation}

\subsection{Approach to the steady state}

In terms of the steady state distribution $q_\infty$, the long time asymptotics of \eqref{totaloil} are 
\begin{align}
q(x,t) & \approx q_{\infty}(x) + f(x) e^{-\lambda_1 t}, \nonumber \\
 \frac{\partial}{\partial x}\left[D(x) \frac{\partial f}{\partial x} - u_{e}(x) f\right] & = -\lambda_1 f
\label{final-asymptotics}
\end{align}
$-\lambda_1$ is the largest negative eigenvalue of the above operator with no-flux boundary conditions at $x=0$ and $L$, and $f$ is a corresponding eigenfunction. Eqs.~\eqref{initial-asymptotics}  and \eqref{final-asymptotics} give {\em finite dimensional} approximations to the short and long time behavior of the solutions. We can obtain a uniformly valid approximation by matching  the two descriptions.  

We need to switch from the short time to the long time description once the pulse (Eq.~\eqref{initial-asymptotics}) is close enough to the shore that it feels the effect of the no-flux boundary condition. We thus distinguish short and long times by comparing $\mu^2(t)/\sigma^2(t)$ with a fixed  threshold $a$, which we set at 4 in the rest of the paper (See discussion below). This corresponds to changing from the initial to the final asymptotics when $2.5\%$ of the total oil has ``beached" and $84\%$ of the oil is within $3 \sigma(t)$ from the shore. At the time $t_e$ determined by $\mu(t_e) = \sqrt{a} \sigma(t_e)$ we switch from using \eqref{initial-asymptotics} to \eqref{final-asymptotics} to describe $q(x,t)$, {\em i.e.}, for $t > t_e$
\begin{align}
q(x,t) \approx & q_\infty + \left[ \sqrt{\frac{2}{\pi \sigma^2(t_e)}} \frac{1}{1+\mathrm{Erf}(\sqrt{a/2})} \right.\nonumber \\ 
& \cdot \left. \exp\left[ - \frac{(x-\mu(t_e))^2}{2 \sigma^2(t_e)}\right]- q_\infty \right] e^{-\lambda_1(t-t_e)},
\label{matched}
\end{align}
where we have normalized the Gaussian to have unit mass in $(0,\infty)$ and  $\lambda_1$ is determined numerically by discretizing the operator in \eqref{final-asymptotics}. 

\subsection{Patching asymptotic solutions}

Equations \eqref{initial-asymptotics} and \eqref{matched} together give a (piecewise defined) composite solution which we denote as the {\em reduced} or {\em ODE model} to contrast with the the PDE model in section~\ref{sec:pde}. A natural question is the dependence of the matched solution \eqref{matched} on the (somewhat arbitrary) choice of threshold $a$.  
Insofar as there is an overlap region where both \eqref{initial-asymptotics} and \eqref{final-asymptotics} give good approximations to the true solution, the composite solution is  insensitive to the precise choice of $a$ and gives a good approximation to the true solution from the PDE model for all time (see \cite{benderorszag}). We have also verified this numerically by taking $a =1$ and $a = 9$, which give similar results to taking $a=4$.

 We get the following predictions for the dynamics of pulse solutions from the ODE model. For small times, the Gaussian pulse solution has an maximum away from the shore $x=0$ for $q,s$ and $b$. For long times, the maximum for $q$ (respectively $s$ and $b$) depends on the location of the maximum of the steady state for $q$ (respectively $s$ and $b$), which in turn is determined by the sign of the effective velocity $u_e$ in \eqref{totaloil} (and $\gamma$ and $\xi$). If the steady solution has a maximum away from the shore, then the maximum of $q$ (resp. $s$ and $b$) is away from the shore for all time.  This indicates  {\em nearshore stickiness}. Conversely if the steady solution has maximum at $x =0$, the matched solution will have two local maxima, one away from the shore for the Gaussian pulse, one at the boundary from the steady state. Further the {\em global} maximum will jump instantaneously to the shore at a critical time when the values at the two local maxima are equal.  

Finally, the steady solution $q_\infty$ has a maximum that is significantly away from the shore, leading to nearshore stickiness, provided that $\beta \gtrsim O(1)$  (as shown in \eqref{beta}) and $D(x) > D_{threshold}$ in the breakzone (as shown in \eqref{threshold}). These criteria are robust and depend on the mixing layer depth $P$ and the bottom topography $H_0,m$ and the enhanced diffusivity $D_L$ in the breakzone, but are insensitive to the details of the vertical mixing and in particular to the specific choice for the mixing time $\tau(x)$.

\section{Model Outcomes} \label{sec:results}

In this section we present, compare, and discuss results from numerically solving the PDE model  (Section~\ref{sec:pde}) and the ODE model (Section~\ref{sec:ode}). The PDE model was numerically integrated using a Crank-Nicholson method with centered differencing for the diffusion terms and upwind differencing for the advection terms, while the ODE model was solved using {\tt ode45} of matlab. Table 1 summarizes the values of the parameters used in the illustrative examples
that follow. 
\begin{table*}
  \begin{center}
  \caption{Parameter values for the example depicted in Figures
\ref{velo}-\ref{pl2}.}
\begin{tabular}{||ccc|ccc|ccc||}
\hline
 & units & value &    & units & value &   & units & value \\
\hline 
$H_0$ &  m& 1.2 &  $H_{\infty}$ & m  & 20  & $X$ & m &  1000\\ 
 $L$ & m & 200 & $P$ & m & 1 -- 6& $ w$ & m & 20 \\
 $D_{eddy}$ & m$^2$/s &  0.05 & $D_{L}$ & m$^2$/s & 1.6 &     ${\cal U}^{St}$ & m/s&  $0.0203$\\  
 \hline
\end{tabular}
\label{tab1}
\end{center}
\end{table*}
Before embarking on results, we can use $\beta$ and $\delta$ to predict, approximately, 
what conditions are required for stalling outside of the break zone, if we accept the parameter values in Table 1. From (\ref{beta}), we see that for stalling to occur near the edge of the breakzone, we require a $P \gtrsim 3.76$m. From (\ref{delta}), and for $P \approx 3.76$ m,  we can ask whether $D_L=1.6$ m$^2$/s is above the $D_{theshold}$ required to see stalling.
We find, approximately, that $D_{threshold} = 4\frac{\gamma}{1-\gamma}$.
Hence, $\gamma \lesssim 0.29$, for $D_{threshold} \lesssim 1.6$ m$^2$/s.

Figure \ref{veloa} displays the velocity $u(x,t)$ at $x=X$. Superimposed is a dashed line indicating $P=1$.  Figure \ref{velob} depicts the diffusivity $D(x)$, used in the calculations. 
 The effective velocity $v(x)$ in (\ref{altoilb}) is shown in Figure \ref{veloc}, for several $P$.
 For various choices of $P$ and $\gamma$, we  describe the evolution from an initial condition corresponding to a symmetric oil slick on the surface, namely
 \begin{align*}
 s(x,t=0) & =  \exp (-0.001 (x-500)^2 )/\sqrt{1000 \pi}\\
   b(x,t=0)& = 0.
 \end{align*}
  
\begin{figure*}
\begin{center}
\subfigure[]
{
\includegraphics[width=3.2in]{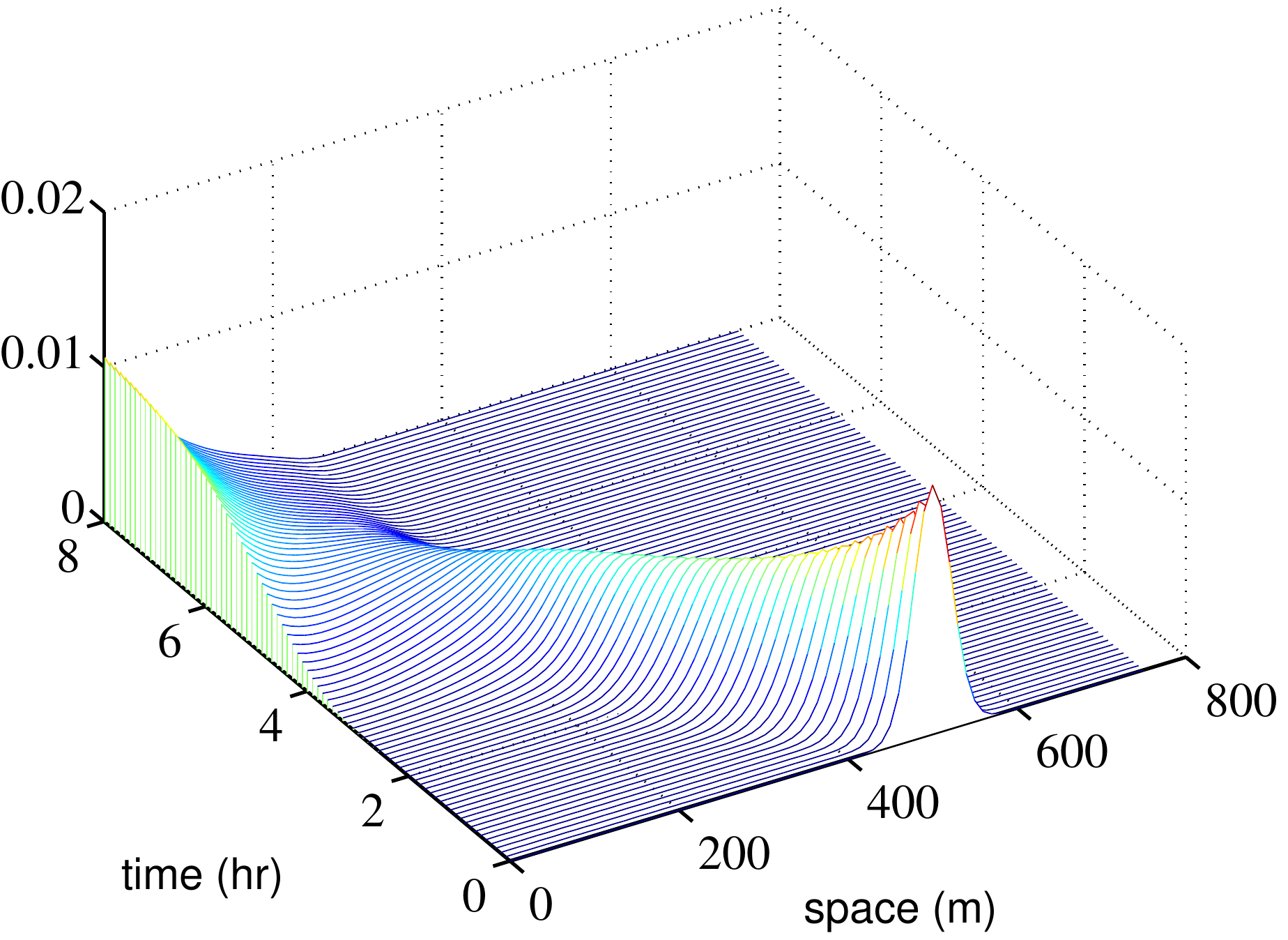}
\label{sta}
}
\subfigure[]
{
\includegraphics[width=2.8in]{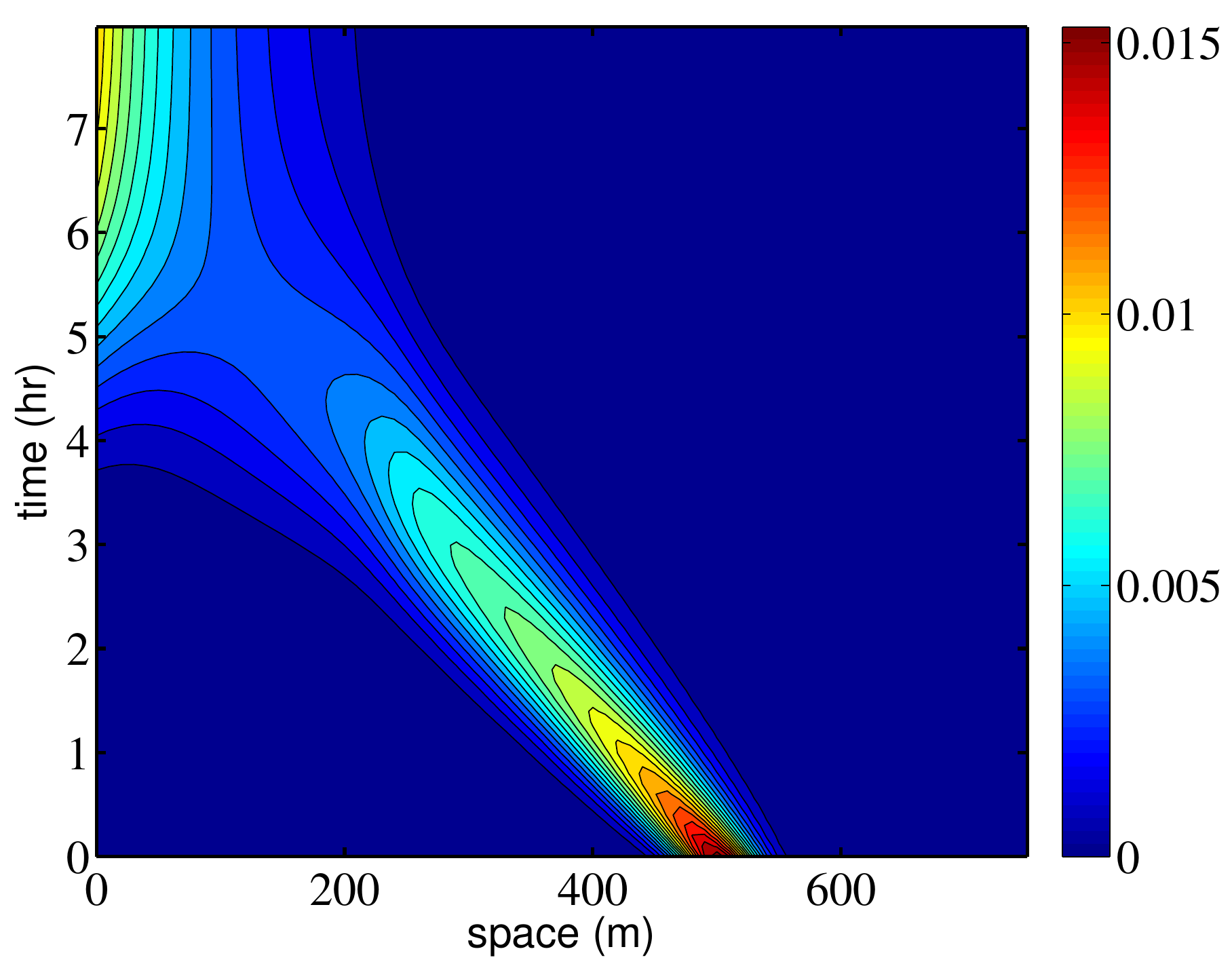}
\label{stb}
}
\caption{\label{st} Evolution of the \subref{sta} oil slick $s(x,t)$;  \subref{stb} contours of $s(x,t)$,
 $P=1$, $\gamma=0.9$. For this case, $\delta = 0.17$, $\beta=-0.05$.}
\end{center}
\end{figure*}

 \noindent{\bf Case I:} $P=1, \gamma=0.9$.  Since $\gamma$ is high most of the oil is in the slick, rather than the interior.   The space-time evolution of the $s(x,t)$ field is shown in Figure \ref{sta}, 
and Figure~ \ref{stb} depicts the contours of $s(x,t)$.  The dynamics of the oil slick and the interior oil are phase-locked because the vertical mixing is very rapid on the scale of the horizontal transport.  The $b(x,t)$ field takes up some of the oil 
and travels at the same rate as the surface oil $s(x,t)$. 
\begin{figure*}
\begin{center}
\subfigure[]
{
\includegraphics[width=0.45 \textwidth]{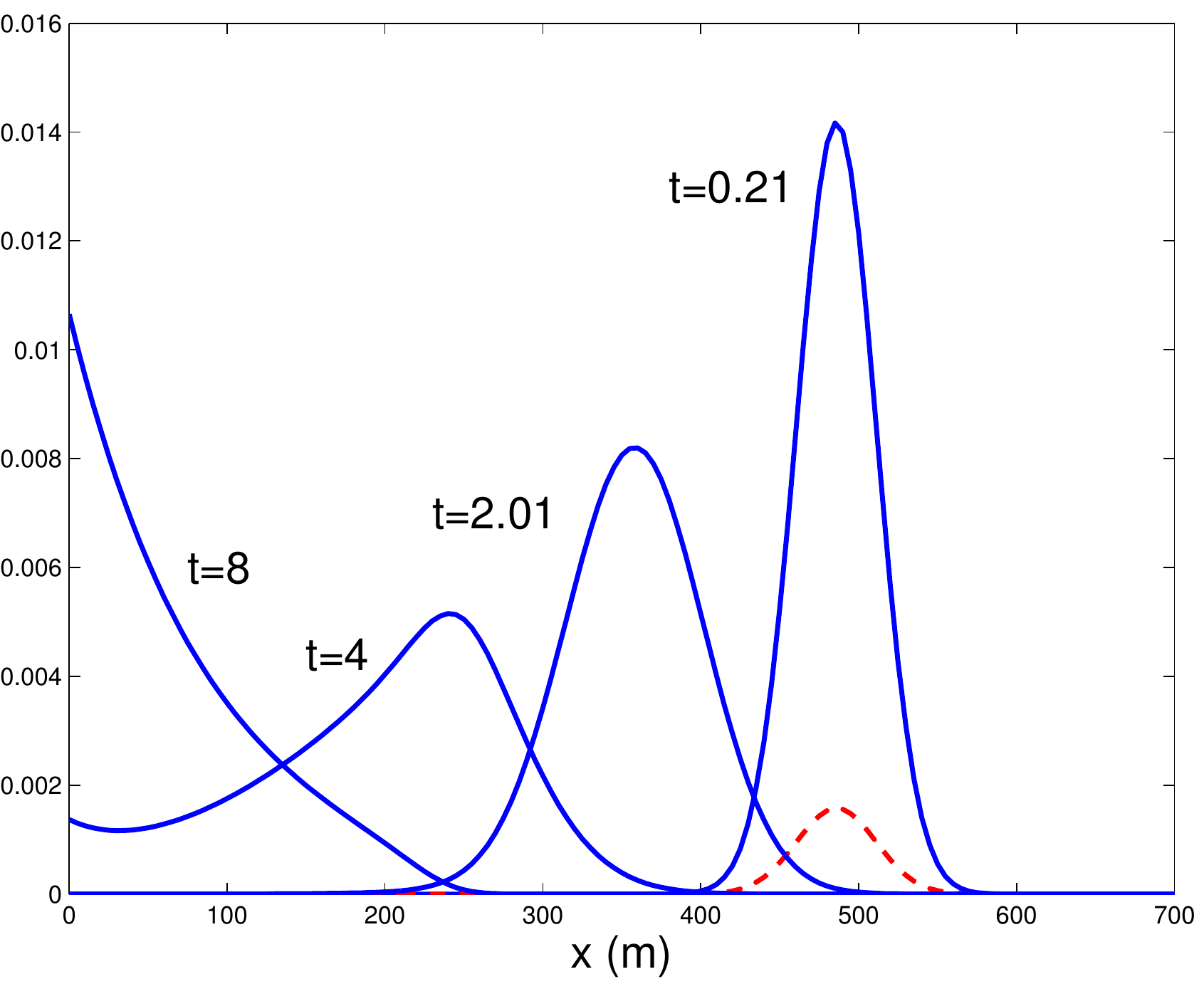}
\label{evoa}
}
\subfigure[]
{
\includegraphics[width=0.45 \textwidth]{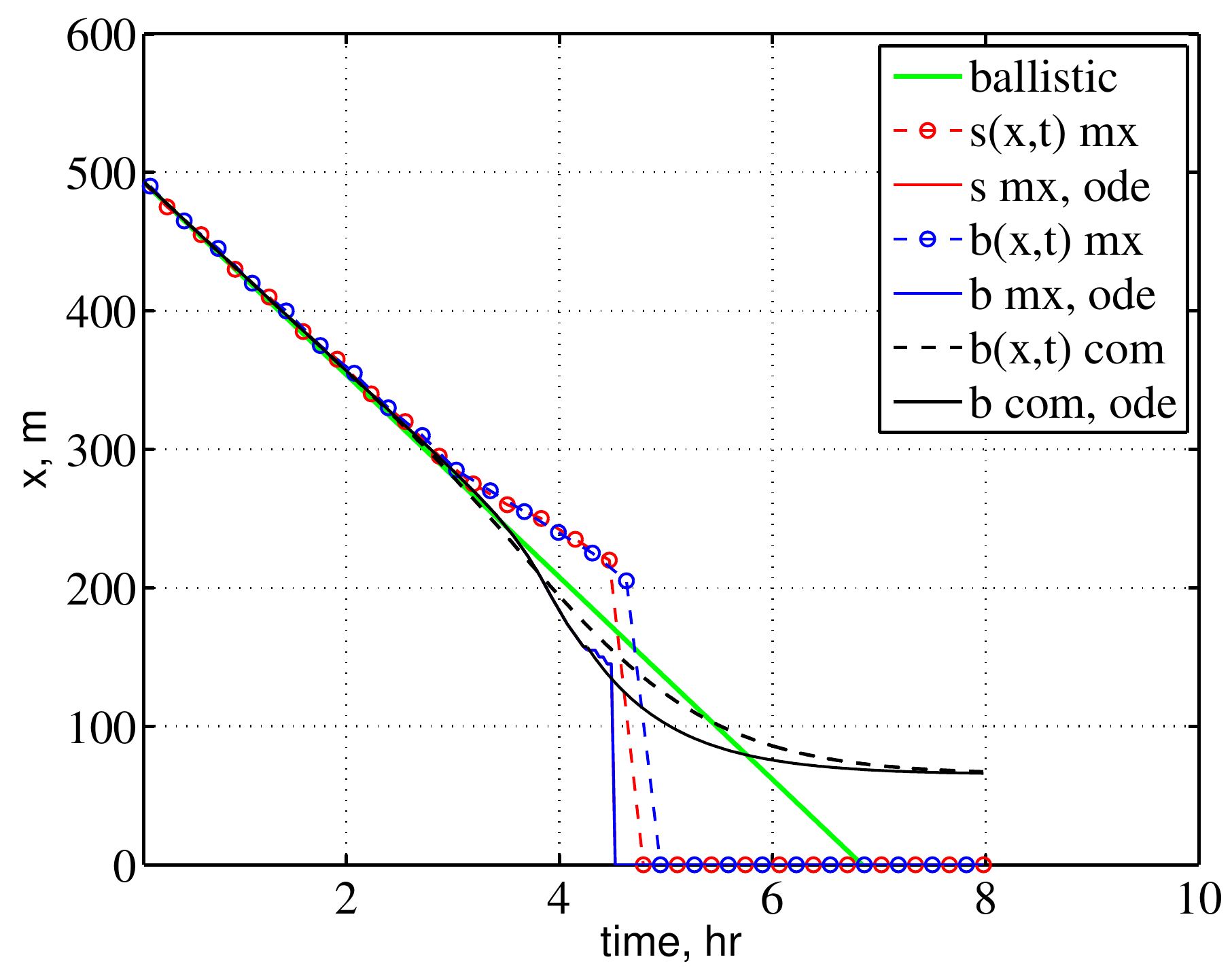}
\label{evob}
}
\caption{\label{evo} \subref{evoa} Evolution of oil slick and interior oil, at four different times.
$P=1$, $\gamma=0.9$.  The oil slick (solid), which is initially
symmetric about its center of mass, changes shape dramatically as it approaches the shore. 
Dashed, $b$ at $t=0.21$ hrs is shown, for scale. Subsequent $b(x,t)$ are similar in shape to $s(x,t)$.
 \subref{evob} The maximum (mx) of the oil slick  as well as its center of mass (com) slow down slightly outside of the nearshore, compared to the ballistic trajectory. The maxima of $b$ and $s$ are
 tracing nearly identical trajectories. }
\end{center}
\end{figure*}
  
  Figure \ref{evoa} illustrates
 the  slick $s$, at four different times. For short times, the pulse is symmetric. For longer times, the pulse loses its symmetry, broadening more toward the beach due to a combination of the effective advection velocity $u_e$, the enhanced turbulent diffusion $D$ in the near shore and  the no-flux boundary condition. Figure \ref{evo} also includes the ``ballistic" 
 prediction, which is defined as the distance/time relationship given by the advective speed 
 ${\cal U}^{St}$.
  
 Figure \ref{evob} shows the track traced out by the maximum of $s(x,t)$  (mx) and the center of mass (com).
 For comparison we superimpose  a   hypothetical  {\it ballistic} trajectory, for oil traveling at speed ${\cal U}^{St}$. A  slowdown of the pulse is evident. It begins to happen  before the  oil reaches the edge of the turbulent nearshore, which here spans $x=[0,200]$ m.   Since $P<H(x)$, $u_B(x), v(x)$ and $u_e(x)$ are never zero ({\it cf.}, Figure \ref{veloc}). The steady state solutions for $s$ and $b$ have their maxima at $x=0$, and as one would expect the maxima for both $s$ and $b$ jump instantaneously from the interior to the boundary. The center of mass of the oil slick, should approach the center of mass of the steady distribution, and thus stays away from the shore. We also plot the ODE model trajectories of the maximum and the center of mass. The agreement between the PDE and ODE trajectories is excellent.
  
\begin{figure*}
\begin{center}
\subfigure[]
{
\includegraphics[width=0.3 \textwidth]{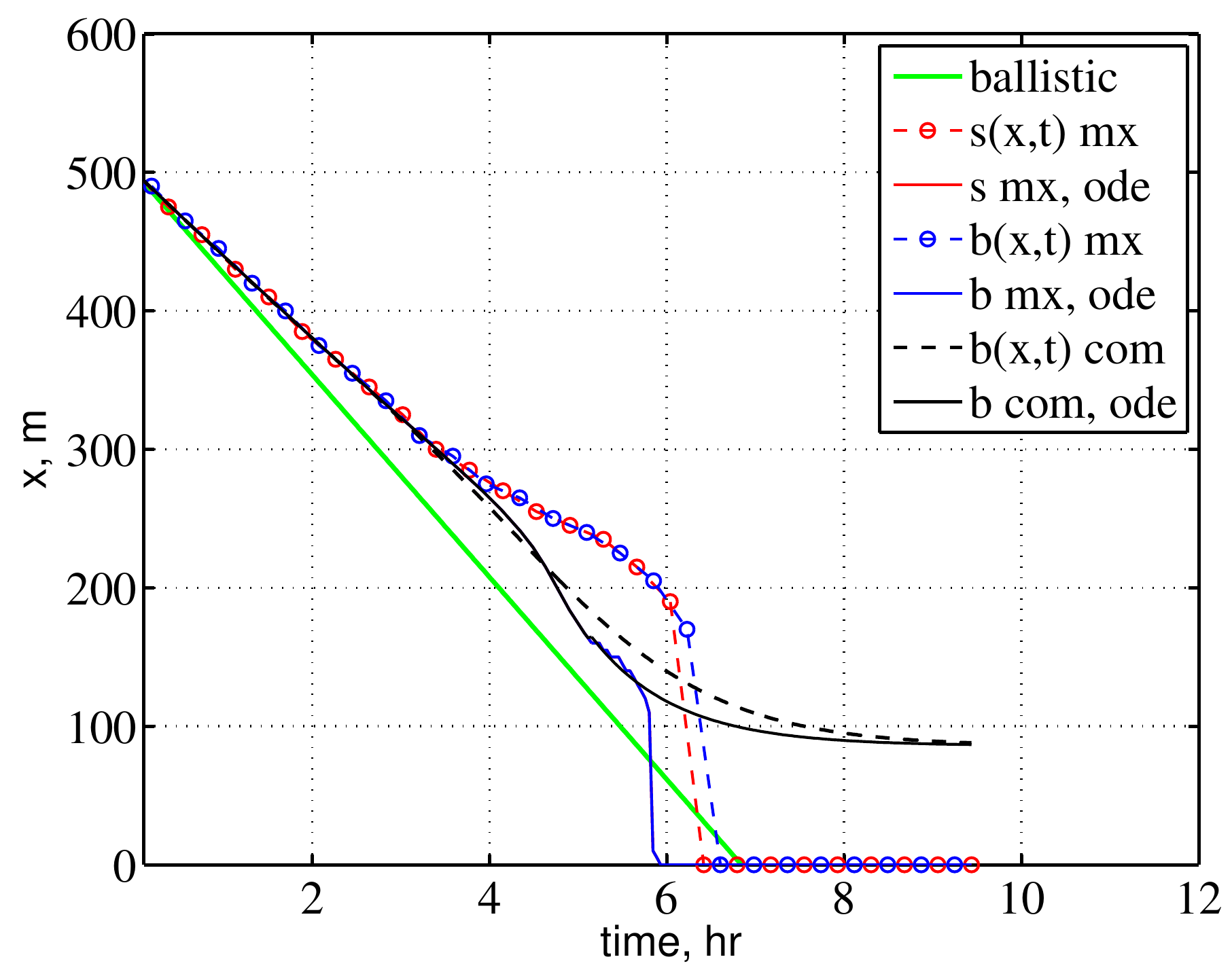}
\label{evogamma-a}
}
\subfigure[]
{
\includegraphics[width=0.3 \textwidth]{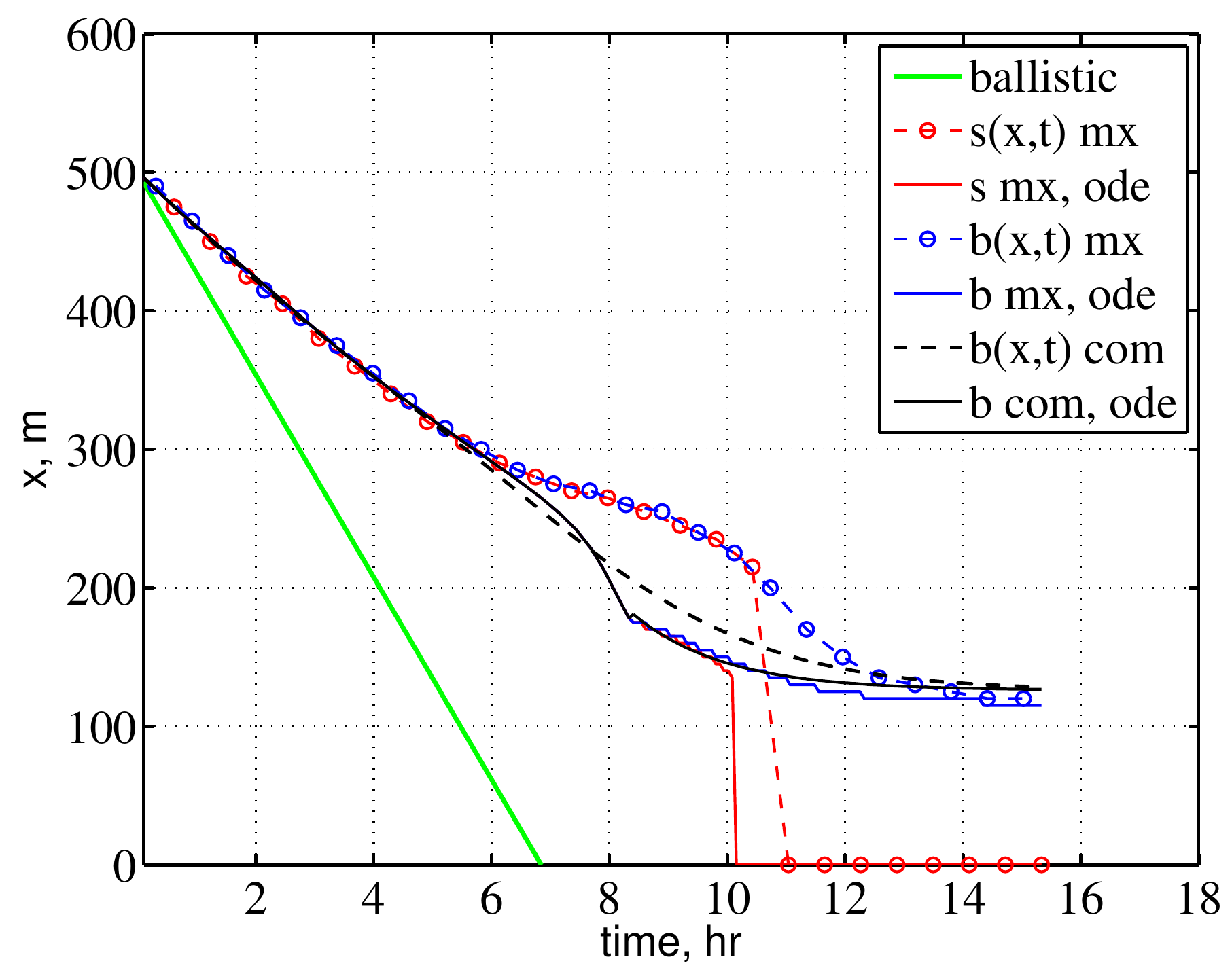}
\label{evogamma-b}
}
\subfigure[]
{
\includegraphics[width=0.3 \textwidth]{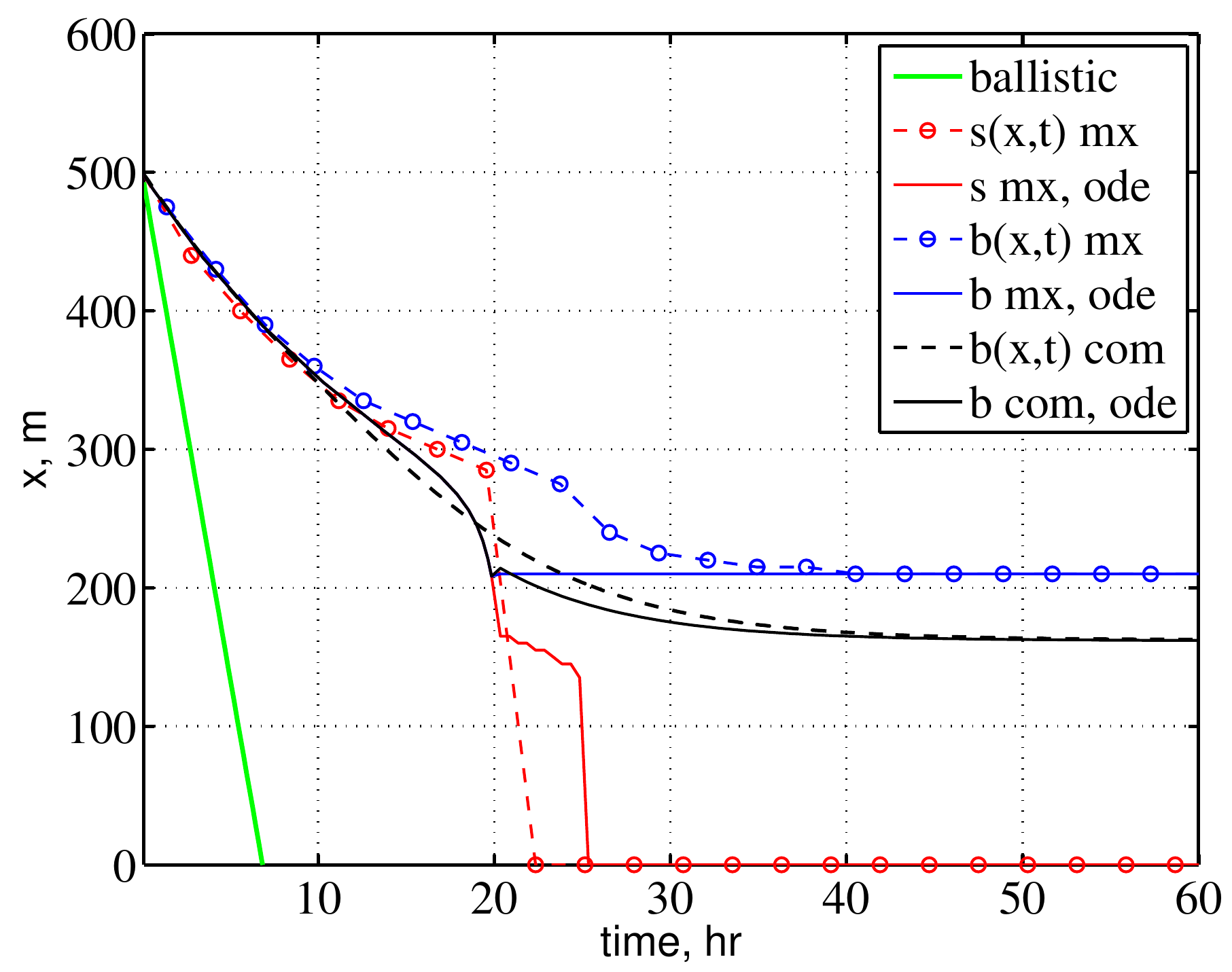}
\label{evogamma-c}
}
\caption{\label{evogamma}Trajectories of maxima (mx)  and centers of mass (com), $\gamma=0.1$, \subref{evogamma-a} $P=1$, \subref{evogamma-b} $P=3$, \subref{evogamma-c} $P=6$.  As is evident as well in Figure~\ref{evo}, the ODE model captures the 
dynamics of the full model very well.}
\end{center}
\end{figure*}

\begin{figure*}
\begin{center}
\subfigure[]
{
\includegraphics[width=0.3 \textwidth]{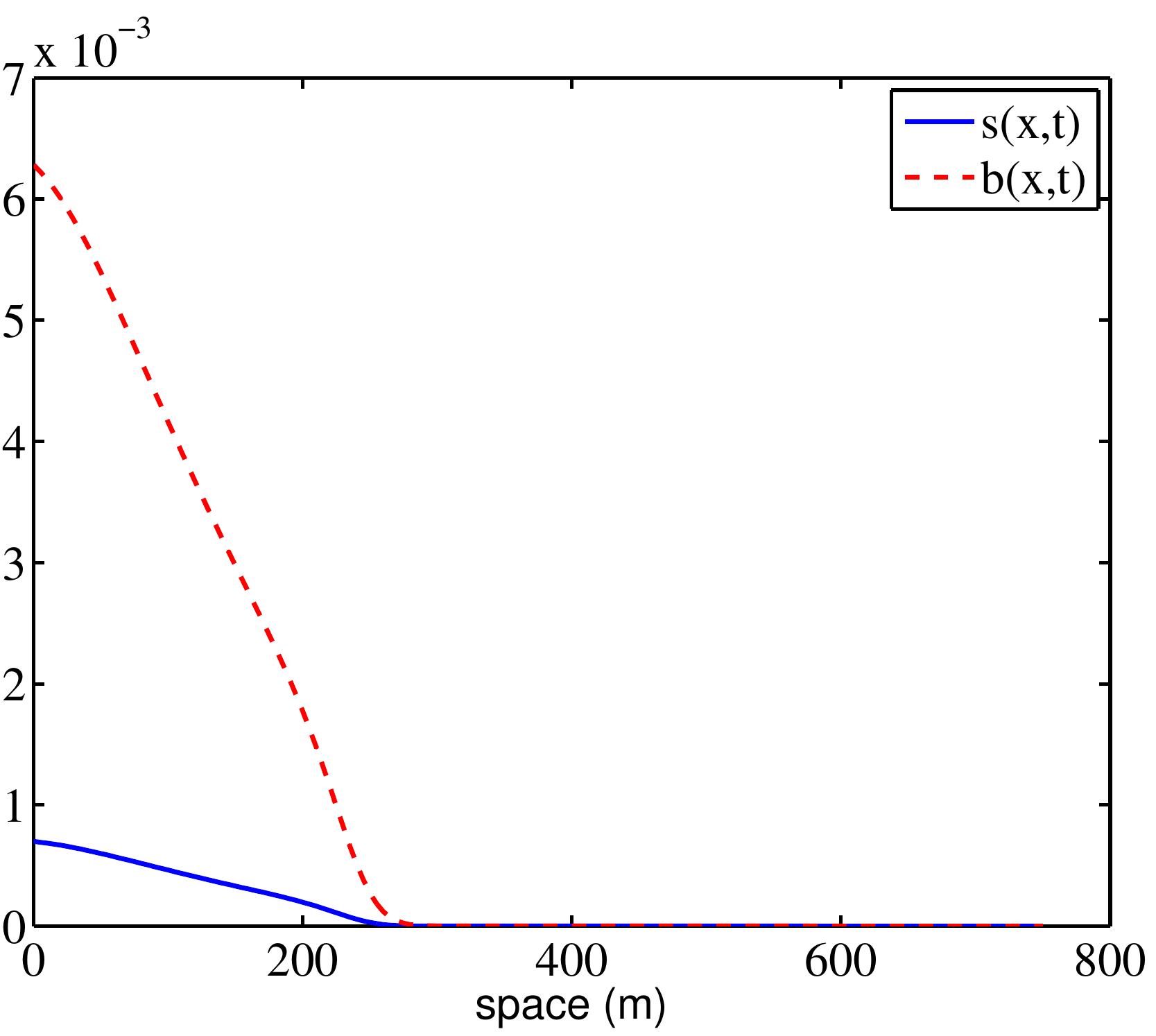}
\label{pl2-a}
}
\subfigure[]
{
\includegraphics[width=0.3 \textwidth]{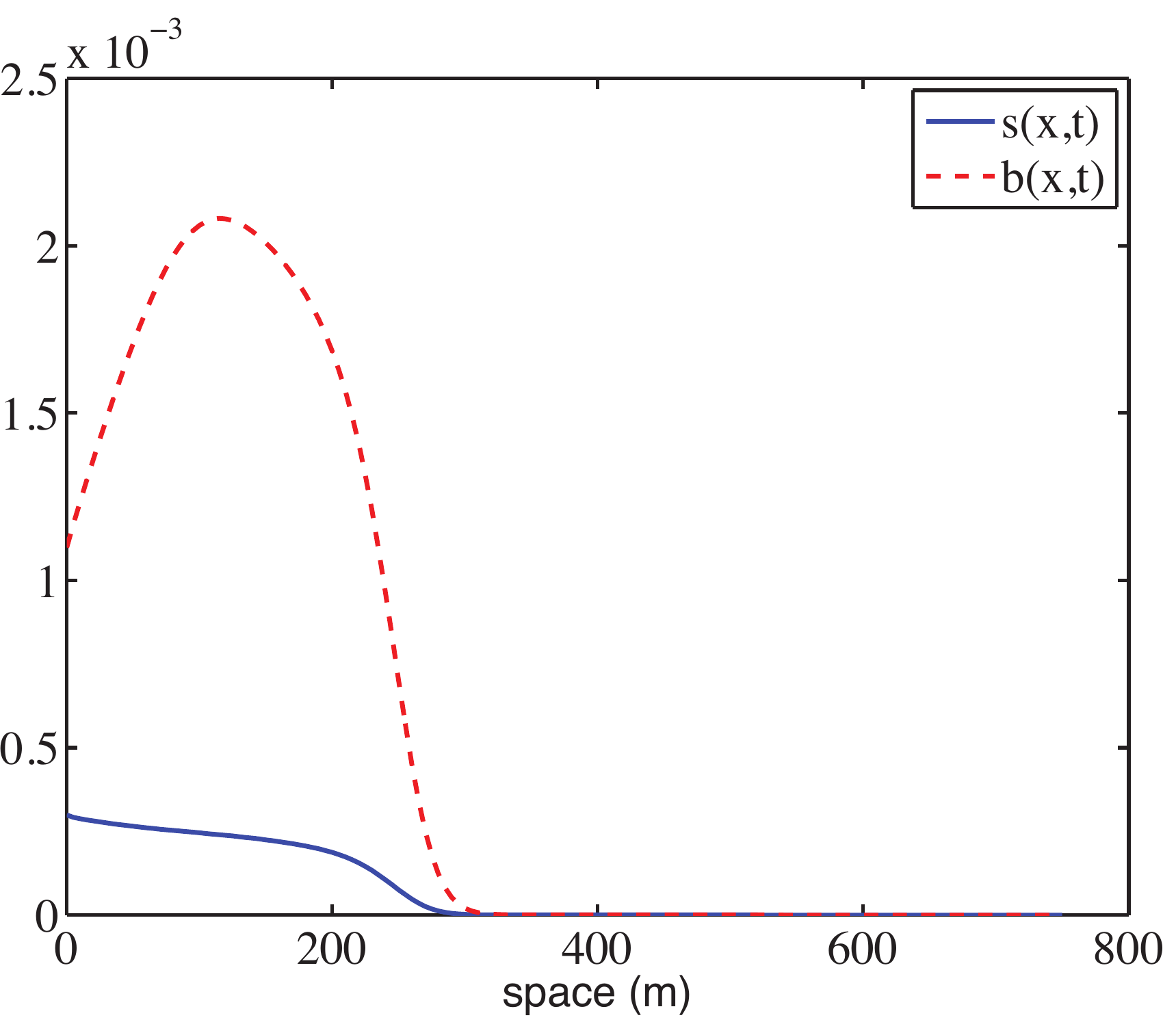}
\label{pl2-b}
}
\subfigure[]
{
\includegraphics[width=0.3 \textwidth]{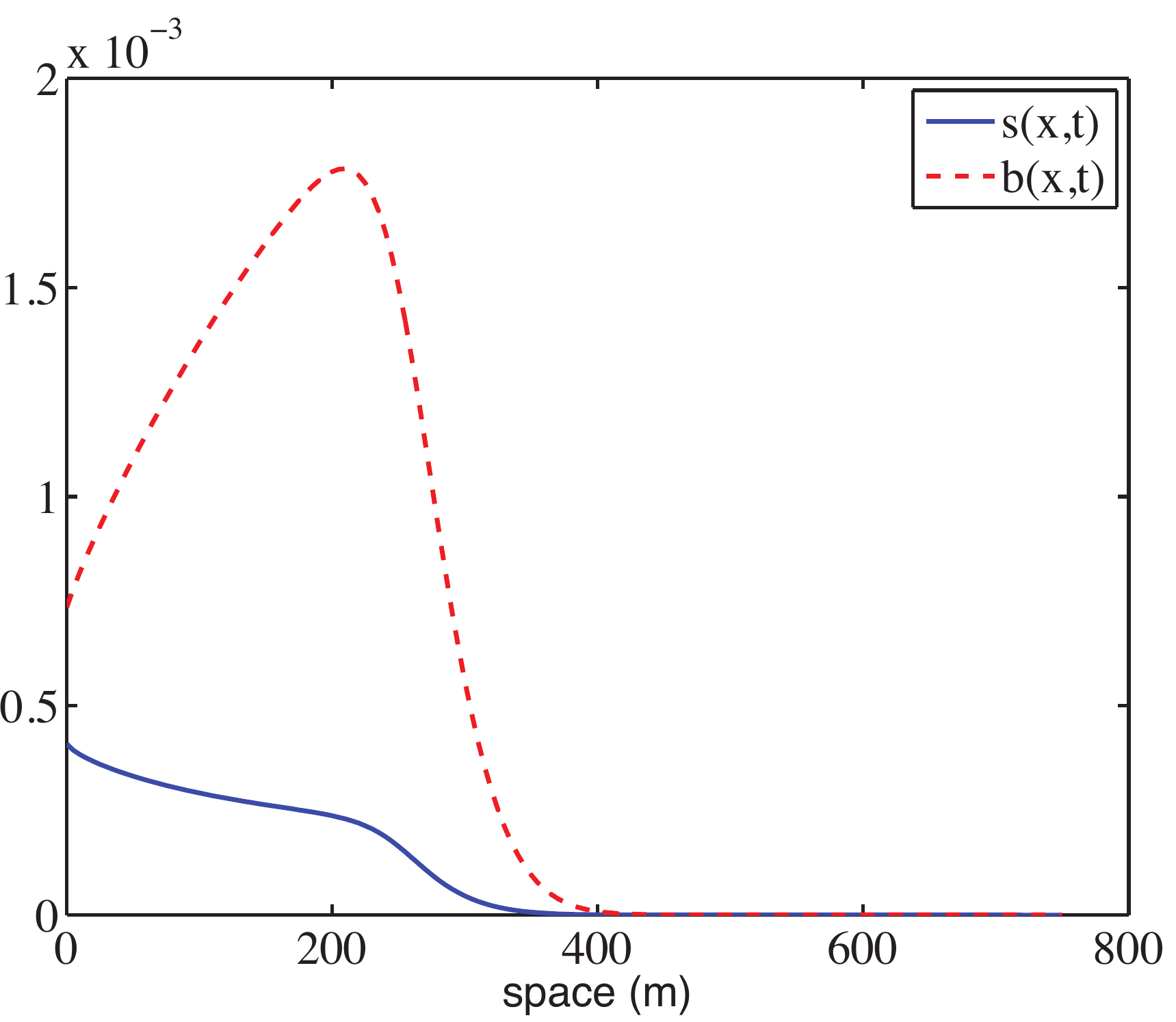}
\label{pl2-c}
}
\caption{\label{pl2} Steady state of $s$ and $b$. $\gamma = 0.1$. \subref{pl2-a} $P=1$, $\delta=13.34$, $\beta = -0.05$;  \subref{pl2-b} $P=3$, $\delta=4.45$, $\beta = 0.48$; \subref{pl2-c} $P=6$, $\delta=7.22$, $\beta = 1.28$. Note that as $P$ is increased to 3 and beyond, the location of the maximum moves away from the shore to the edge of the surf zone.
}
\end{center}
\end{figure*}

\noindent{\bf Case II:} $P=1, \gamma = 0.1$. As $\gamma$ is decreased $u_e$ gets closer to $u_B$ and $v$, which are smaller in magnitude than  ${\cal U}^{St}$. This causes a slowdown of the center of mass of  $q$ compared to the ballistic trajectory. 
(See Figure \ref{evogamma-a}.) 
The oil, which started in the slick, transitions to the interior and the bulk of the oil 
in short order. 
 We note the $v$ and $u_e$ for this case and the previous case are strictly negative, because $P=1$ (See Figure~\ref{veloc}), so the maximum for $s$ and $b$ in the steady state is at $x=0$ (See Figure~\ref{pl2-a}.) Thus we expect the maxima to jump to zero, and this is borne out by simulations from the full model and also the reduced model, as depicted in Figure~\ref{evogamma-a}. 

 \noindent{\bf Cases III and IV:} $\gamma = 0.1, P = 3$ and $P = 6$ respectively. 
 In these cases, $P=H(x)$ at $x= 95$ m (inside the break zone) and $x = 254$ m (outside the break zone), respectively. The effective velocity $v$ is positive
on the shallow side of $P=H$, as shown in Figure~\ref{veloc} and the effective velocity $u_e$ changes sign. The maximum for the steady states of $b$ are now in the interior of the domain, while the maximum for the steady states of $s$ remain at $x = 0$ (See Figures~\ref{pl2-b}-\subref{pl2-c}). We thus predict that the maxima for $s$ will jump instantaneously, but the maxima for $b$ will remain away from the shore. This is indeed the case, for both the full and the reduced models as shown in Figures~\ref{evogamma-b}-\subref{evogamma-c}. 

\begin{figure*}
\begin{center}
\subfigure[]
{
\includegraphics[width=0.45 \textwidth]{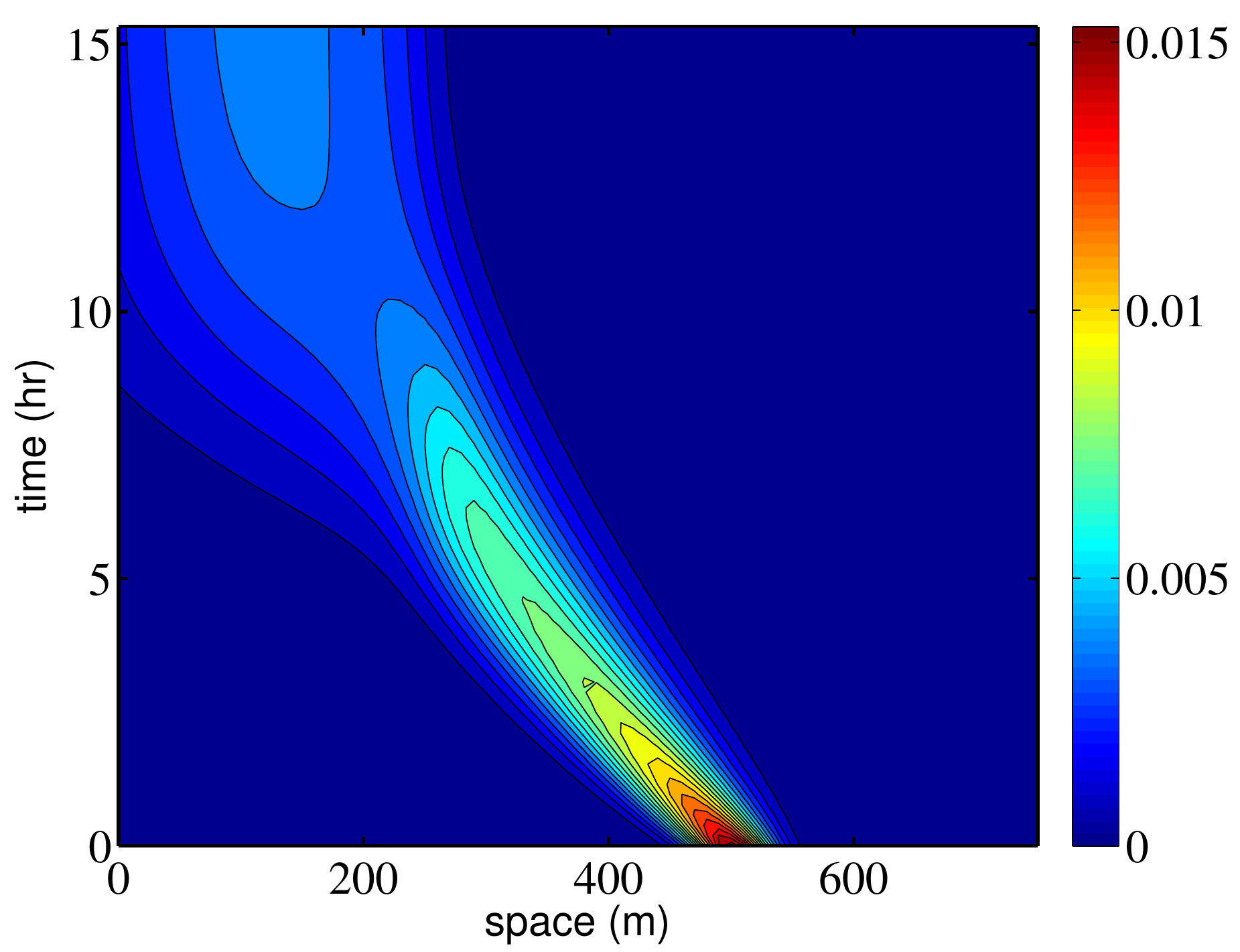}
\label{pl1-a}
}
\subfigure[]
{
\includegraphics[width=0.45 \textwidth]{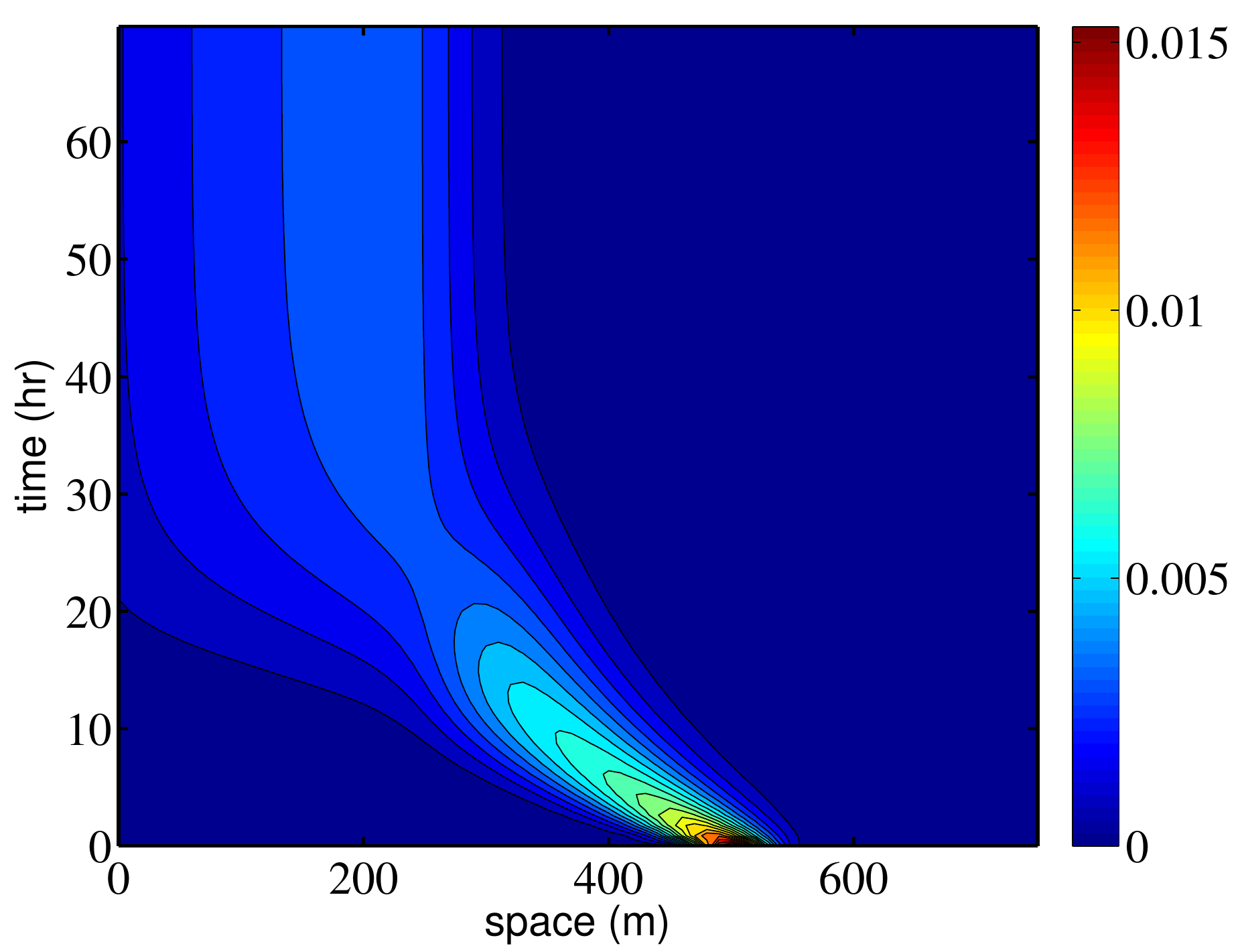}
\label{pl1-b}
}
\caption{\label{pl1} Contours of $b(x,t)$; $\gamma = 0.1$,  \subref{pl1-a} $P=3$,  \subref{pl1-b}  $P=6$. }
\end{center}
\end{figure*}

Since $\gamma=0.1$, the initial pulse of oil in the slick transitions quickly to the interior, and remains there. Also, the maximum of $b$ stays away from the shore, so this is an example of {\em nearshore sticky waters}. The space-time evolution of $b(x,t)$ is shown in Figure~\ref{pl1}, showing the slowing and parking of the tracers in the bulk away from the shore. We wish to draw attention to the agreement between the PDE and the ODE models in Figures~\ref{evogamma-b}-\subref{evogamma-c}, even in situations where the ballistic and actual trajectories are very different.

\section{Conclusions}

The model we propose describes, in a simple way, how buoyant contaminants that are impervious to inertial effects, may slow down as they approach
nearshore environments. Evidence of unusually long residence times for tracers, measured in
breakzone field experiments, is reported from time to time (see for example, \cite{reniers09}). 
The mechanism depends on vertical variations of the tracer density and the cross-shore component of the velocity, 
and thus on buoyant stratification and topographic features,  rather than on blocking or stationary structures in the Lagrangian trajectories induced by  the velocity field (see \cite{lcs}, for details on Lagrangian Coherent Structures). It is also different from the surface-wave deceleration mechanism outside of the surf zone, that is explored in \cite{ohlmann12}.

The phenomenon and the dynamics detailed here should have relevance to the long-term transport of other contaminants and biological material in which inertial effects are negligible. In environmental mitigation problems the timing of landfall of pollutants and the manner in which these make it to land are crucial. 
We hope this simple model motivates efforts to improve 
forecasting circulation models for nearshore flows and transport, with better the modeling of breakzone boundary conditions, and nearshore subscale transport of pollutants.

The proposed model (Section~\ref{sec:pde}) is a set of coupled PDEs and, despite its relative simplicity, it  has multiple, potentially relevant parameters (See Table 1). In order to gain further insight into the dynamics of the PDE model, we derived a finite dimensional reduction in Section~\ref{sec:ode}. We gain significant insight from studying the reduced model, namely we can explain the evolution of the vertical stratification of the oil density, the evolution of the oil density from symmetric to shoreward skewed distributions, the eventual fate of the maximum and the center of mass of the oil distribution, both in the surface and in the interior including an explanation of stickiness phenomenon, and a prediction for the time and space scales for the complex dynamics of the oil distribution. 

Most importantly, the reduced model demonstrates that nearshore stickiness is a robust phenomenon,  and helps identify the key parameters (or combinations thereof) which govern the gross behavior of the system. The determining factors are the two dimensionless quantities
 $$
 \beta = \left(\frac{P-H_0}{H_\infty-H_0} \right) \frac{X}{L}, \qquad \delta =  \frac{(1-\gamma) D_L (H_\infty - H_0)}{\gamma P |{\cal U}^{St}| X}.
 $$
  These quantities indicate the outcomes  would change if the 
 topographical parameters or the  magnitude of the ${\cal U}^{St}$, $P$, $\gamma$, and/or $D_L$ are changed.
 A general conclusion from our analysis is that increasing $\beta$ and/or increasing $\delta$ (equivalently decreasing $\gamma$)  leads to more stickiness, {\it i.e}, the oil approaches the shore more slowly and stalls near the break zone for much longer. 

The parameter $\beta$ depends on the topography of the nearshore, the sea conditions and the nature of the pollutant, as all of these influence $P$. Locations where the topography drops steeply even near the shore, such as in the Southern California Bight, will have small $\beta$ and are thus less ``sticky." Conversely a typical situation in the Gulf of Mexico or on the mid and southern portions of the
Atlantic  US coast would have $P$  large  when compared to $H$, even at considerable distances from the shore, i.e. $\beta$ large, so the nearshore is considerably more ``sticky." The parameter $\delta$ depends on the enhanced turbulent diffusivity $D_L$ in the breakzone as well as the parameter $\gamma$ which governs the fraction of the oil which is in the slick. Increasing the turbulent eddy diffusivity or decreasing the fraction of the total oil in the slick both lead to increased stickiness.

The models we have developed have the virtue of simplicity and being analytically tractable, but they lack many of the ingredients of more realistic physical models. In particular we do not include inertial effects, and the transient setup and setdown process (see \citet{lentzfewings}). One expects the mixed layer to have a diffuse boundary unlike our simple model with a sharply discontinuous density at a constant layer depth $P$. The mixed layer can depend on the sea state and the oil or pollutant constituents. For instance, the mixed layer depth and the concentration of the oil in suspension will tend to increase with  Langmuir turbulence and  wave breaking activity. We use a kinematically specified flow with a parabolic profile, but a more realistic model should include a dynamical model for the ocean with a realistic vertical mean Lagrangian velocity structure, at least close to the shore, as well as a careful treatment of the form-stress for the 
coupling of wind effects, which are recognized as very important in the case of surface oil slick dynamics (see \cite{windoil}). Finally, we have a simplified description of buoyancy and vertical mixing effects. These are important issues which we consider in a sequel to this paper. While these effects are relevant, we believe they will not change the basic conclusions we have drawn from the simplified model, and it gives a robust description of the mechanisms which can lead to the the observed apparent stalling of incoming pollutants approaching a shore.

\section*{Acknowledgements}

We received funding from GoMRI/BP and from NSF DMS grant 1109856. JMR wishes to thank the Statistics and Applied Mathematical Sciences Institute,
an NSF funded institute, in which some of this research was done. JMR
also thanks the J. Tinsley Oden Fellowship program at the University
of Texas for its support. SCV was also supported in part by the NSF DMS grant 0807501. Prof. J. C. McWilliams is acknowledged
 for bringing to our attention the problem of 
sticky waters in the Great Barrier Reef, which lead us to ask whether sticky waters
 occur in the nearshore setting. We are also very grateful to Prof. F. Feddersen for
 sharing with us his expertise on dispersive processes in the surfzone. His suggestions
 considerably improved this paper.

\begin{table*}
 \begin{center}
  \caption{Table of Symbols.}
\scalebox{0.85}{
\begin{tabular}{||ccc||}
\hline
name &symbol & units \\
\hline 
transverse coordinate &$ {\bf x}=(x,y)$ & m \\
depth coordinate &$z$ & m \\
cross-shore coordinate & $x$ & m \\
alongshore coordinate & $y$ & m \\
time   & $t$ & s\\
upward unit vector & ${\bf \hat z}$ & -\\
cross-shore unit vector & ${\bf \hat x}$ & - \\
gravity & $g$ & m/s$^2$ \\
exchange relaxation time & $\tau$ & s \\
minimum depth& $H_0$ &  m \\
 maximum depth & $H_{\infty}$ & m  \\
 topographic extent& $X$ & m \\
surf zone extent& $L$ & m \\
 mixing layer thickness & $P$ & m \\
  transition width & $ w$ & m \\
  bottom gradient & $m$ & m/m \\
  bottom topography & $H(x)$ & m \\
 mean (current) sea elevation &$\zeta^c = \zeta + \hat \zeta$& m \\
 quasi-static sea elevation correction & $\hat \zeta$ & m \\
 sea elevation & $\zeta$ & m \\
 total water column & ${\cal H}=H(x,t)+\zeta^c$ & m \\
 3D velocity & ${\bf U}({\bf x},z,t)=(U,V,W)$ & m/s \\
 depth-averaged Eulerian velocity & ${\bf v}^c({\bf x},t)=(u^c,v^c)$ & m/s \\
depth-averaged Stokes drift velocity &  ${\bf u}^{St}({\bf x},t)=(u^{St},v^{St})$ & m/s\\  
cross-shore Stokes drift, at the surface & ${\cal U}^{St}$& m/s\\
slick advective velocity & $u_S$ & m/s\\
bulk advective velocity & $u_B(x)$ & m/s\\
effective bulk oil velocity &$v(x) = u_B(x) + D(x) \frac{1}{\xi(x)} \frac{d \xi(x)}{dx}$ & m/s\\
oil mixed depth & $\xi \approx \min(H(x),P)$ & m \\
average oil velocity (ODE) & $u_e$ & m/s \\
horizontal oil dispersion & $D=D_{eddy} + {\cal S}(x) D_L$ & m$^2$/s \\
 oil dispersion transition sigmoid  & ${\cal S}=\left(1 +\exp(\frac{x-L}{w})\right)^{-1}$ & - \\
horizontal dispersion, surf  & $D_L$ & m$^2$/s \\
horizontal dispersion, far-field &  $D_{eddy}$ & m$^2$/s \\
horizontal threshold dispersion & $D_{threshold}$ & m$^2$/s \\
slick thickness & $s(x,t)$ & m \\
bulk oil thickness & $b(x,t)$ & m \\
bulk oil volume fraction & $B(x,t)$ & -\\
vertical vorticity & $\chi$ & s$^{-1}$ \\
wave action & ${\cal A}$ &  Kg/s\\
loss term in the action equation & $N_{{\cal A}}$ & Kg/s$^2$ \\
wave height & $A$ & m \\
relative group velocity & ${\bf C_G}$ & m/s \\
wavenumber vector & ${\bf k}$ & m$^{-1}$ \\
wavenumber magnitude & $k$ & m$^{-1}$ \\
wave frequency & $\Sigma$ & rad/s \\
relative wave frequency & $\omega$ & rad/s \\
tracer & $\theta$ & tracer units \\
tracer diffusion  & $N_\theta$ & tracer units/s \\
momentum source/sinks/dissipation & ${\bf N}$ & m$^4$/s$^2$/Kg \\
average oil distribution (ODE) & $q$ & m\\
longtime  oil distribution (ODE) & $q_\infty$ & m\\
smallest (in magnitude) non-zero eigenvalue (PDE) & $\lambda_1$ & s$^{-1}$\\
eigenfunction associated with $\lambda_1$  (PDE) & $f(x)$ & m\\
switching time (ODE) & $t_e$ &  s\\
proxy for peak of oil distribution (ODE) & $ \mu(t)$ & m \\
variance of oil distribution (ODE)  & $\sigma^2(t)$ & m \\
ratio of mean-square to variance at $t_e$  & $a$ & -\\
ratio of surface oil to total oil& $\gamma$ & - \\
stalling parameter & $\beta$ & - \\
ratio of dispersion & $\delta=D_L/D_{threshold}$ & - \\
proportionality constant & $C$ &  m\\
 \hline
\end{tabular}
}
\label{tab2}
\end{center}
\end{table*}

\end{document}